\documentclass[prb,twocolumn,a4paper,aps,superscriptaddress,amsmath,amssymb]{revtex4}

\usepackage{graphicx}
\usepackage{dcolumn}
\usepackage{bm}
\usepackage{ulem}
\usepackage{epsfig}
\usepackage{float}
\usepackage{color}
\usepackage{textcomp}
\usepackage[lofdepth,lotdepth,caption=false]{subfig}
\usepackage[breaklinks=true,colorlinks,citecolor=blue,linkcolor=blue,urlcolor=blue]{hyperref}
\usepackage{braket}
\usepackage{soul}

\definecolor{sapphire}{rgb}{0.03, 0.03, 0.41}
\DeclareMathOperator{\sech}{sech}
\DeclareMathOperator{\csch}{csch}

\begin{document}
\title{Thermal drag in electronic conductors}

\author{Bibek Bhandari}
\affiliation{NEST, Scuola Normale Superiore and Istituto Nanoscienze-CNR, I-56127 Pisa, Italy}

\author{Giuliano Chiriac\`o}
\affiliation{Department of Physics, Columbia University, New York, New York 10027, United States}

\author{Paolo A. Erdman}
\affiliation{NEST, Scuola Normale Superiore and Istituto Nanoscienze-CNR, I-56127 Pisa, Italy}

\author{Rosario Fazio}
\affiliation{ICTP, Strada Costiera 11, I-34151 Trieste, Italy}
\affiliation{NEST, Scuola Normale Superiore and Istituto Nanoscienze-CNR, I-56127 Pisa, Italy}

\author{Fabio Taddei}
\affiliation{NEST, Istituto Nanoscienze-CNR and Scuola Normale Superiore, I-56127 Pisa, Italy}

\begin{abstract}
We study the electronic thermal drag in two different Coulomb-coupled systems, the first one composed of two Coulomb blockaded metallic islands and the second one consisting of two parallel quantum wires. The two conductors of each system are electrically isolated and placed in the two circuits (the drive and the drag) of a four-electrode setup. The systems are biased, either by a temperature $\Delta T$ or a voltage $V$ difference, on the drive circuit, while no biases are present on the drag circuit. In the case of a pair of metallic islands we use a master equation approach to determine the general properties of the dragged heat current $I^{\rm (h)}_{\rm drag}$, accounting also for co-tunneling contributions and the presence of large biases. Analytic results are obtained in the sequential tunneling regime for small biases, finding, in particular, that $I^{\rm (h)}_{\rm drag}$ is quadratic in $\Delta T$ or $V$ and non-monotonous as a function of the inter-island coupling. Finally, by replacing one of the electrodes in the drag circuit with a superconductor, we find that heat can be extracted from the other normal electrode. In the case of the two interacting quantum wires, using the Luttinger liquid theory and the bosonization technique, we derive an analytic expression for the thermal trans-resistivity $\rho^{\rm (h)}_{12}$, in the weak-coupling limit and at low temperatures. $\rho^{\rm (h)}_{12}$ turns out to be proportional to the electrical trans-resistivity, in such a way that their ratio (a kind of Wiedemann-Franz law) is proportional to $T^3$. We find that $\rho_{12}^{\rm (h)}$ is proportional to $T$ for  low temperatures and decreases like $1/T$ for intermediate temperatures or like $1/T^3$ for high temperatures. We complete our analyses by performing numerical simulations that confirm the above results and allow to access the strong coupling regime.
\end{abstract}

\maketitle

\section{Introduction}
\label{intro}
Two electrically isolated conductors placed close together can still be coupled via the Coulomb interaction. As a result, when a bias is only applied
to one conductor, electronic currents can be generated in the unbiased one in such a way that a charge current is dragged in this second conductor.
This phenomenon, {\it the Coulomb drag}, arises because the carriers in the two conductors are subject to a ``mutual friction'', i.e. to scattering processes
mediated by the Coulomb interaction between the two conductors, and can exchange momentum and/or energy.
The phenomenon of drag, first proposed in 1977 by Pogrebinskii~\cite{Pogrebinskii1977} in layered conductors, has so far been studied in a large
variety of systems and it is still the subject of an intense research activity (see Ref.~\onlinecite{Narozhny2016} for a recent review).
\begin{figure}[h]
\includegraphics[width=\columnwidth,clip=true]{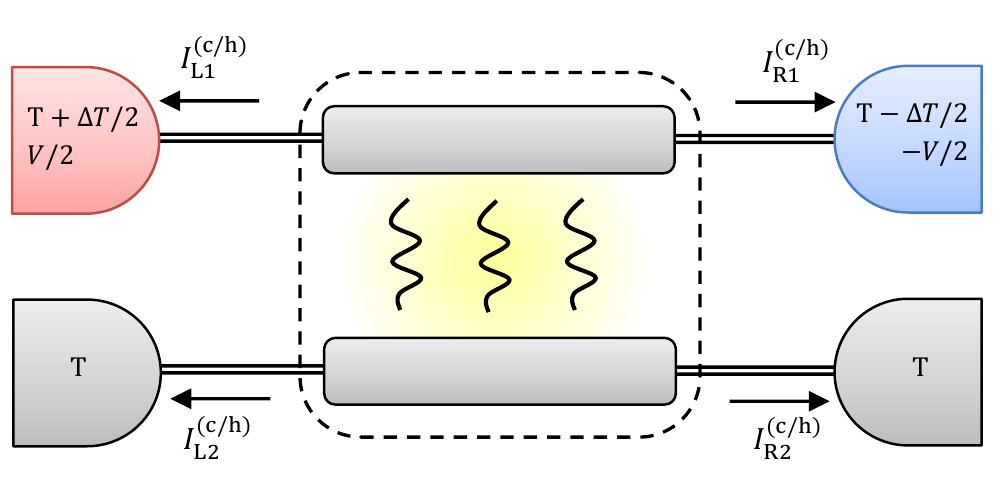}
\caption{Sketch of Coulomb-coupled systems, which consists of an upper (drive) biased circuit, and a lower (drag) unbiased circuit. The conductors are represented as grey rectangles and are attached to two leads each. The two conductors are coupled only through the Coulomb interaction. As indicated by the black arrows, the sign of charge and heat currents is positive when they enter an electrode.}
\label{setup}
\end{figure}

So far most of the attention has been devoted to the effect of drag on the charge current.
In the case of coupled quantum wires the various mechanisms contributing to the  drag were derived  in Refs.~\onlinecite{Pustilnik2003,Flensberg1998}.
They originate from the forward scattering (which is a process with small momentum transfer) or from the backward scattering (where particles are
scattered from one end of the Fermi sphere to the other). Drag measurements in quantum wires have been performed and the predictions made in
Refs.~\onlinecite{Pustilnik2003,Flensberg1998} have been tested, for example, in Refs.~\onlinecite{Debray2001,Debray2000,Laroche2014}.
More recently, the drag of charge between zero-dimensional systems have been theoretically considered for single-level Quantum Dots  (QDs)
in Refs.~\onlinecite{Moldoveanu2009,RSanchez2010,Kaasbjerg2016,Lim2016}.
Experimental investigations in systems composed of two capacitively coupled QDs are reported in Ref.~\onlinecite{Keller2016} (emphasizing
the importance of co-tunneling processes~\cite{Kaasbjerg2016}) and in Refs.~\onlinecite{Bischoff2015,Volk2015} for the case of graphene-based QDs.
The drag of charge in coupled double QDs systems has been also experimentally addressed in Ref.~\onlinecite{Shinkai2009}.
In addition, energy harvesting from thermal and voltage fluctuations in coupled QDs systems attached to three terminals has been considered
theoretically~\cite{RSanchez2011,RSanchez2013,Sothmann2012,Dare2017,Benenti2013} and experimentally~\cite{Hartmann2015,Thierschmann2015,Thierschmann2015-2,Thierschmann2016}.

Another consequence of the Coulomb coupling between two nearby conductors is the fact that a flow of heat can also be induced in the
unbiased conductor. This phenomenon, which is distinguished from the drag of charge that is constrained by the charge conservation within individual
conductors, has been hardly considered in the literature so far~\cite{Whitney2016}.
In the case of metallic islands, heat currents can be induced in the unbiased circuit as a
result of energy transfer, through the capacitive coupling, from the upper island. Such energy transfer has been recently considered for the
implementation of a heat diode~\cite{Ruokola2011}, of a minimal self-contained quantum refrigeration machine~\cite{Venturelli2013}, of a
three-terminal QD refrigerator~\cite{Zhang2015}, of an autonomous Maxwell demon~\cite{Koski2015}, of a Szilard engine~\cite{Koski2014}, of a nanoscale thermocouple heat engine~\cite{Whitney2016}, and for
the study of a correlation-induced in SINIS refrigerator~\cite{RSanchez2017}.
In this paper we will investigate another important case of this kind, {\it thermal drag}.

The setup we consider is represented in Fig.~\ref{setup}. Two mesoscopic conductors, represented by grey rectangles, are coupled through
Coulomb interactions, but cannot exchange electrons. One conductor is contained in the upper (drive) circuit, which is either voltage or thermal biased,
while the other conductor is part of the lower (drag) circuit, which is unbiased. As specified in Fig.~\ref{setup}, the left (right) electrode in the drive circuit
is kept at a voltage $\pm V/2$ and temperature $T\pm \Delta T/2$, while the electrodes in the drag circuit are kept at the same temperature $T$ and at
zero voltage. Our goal is to study the general properties of the heat currents flowing in the drag circuit, $I^{\rm (h)}_{\rm L2}$ and $I^{\rm (h)}_{\rm R2}$, as a
result of energy transfer between upper and lower circuits, due to Coulomb interaction.

We define the drag currents as
\begin{equation}
\label{drag}
I^{\rm (c/h)}_{\rm drag}=\frac{I^{\rm (c/h)}_{\rm L2}-I^{\rm (c/h)}_{\rm R2}}{2} ,
\end{equation}
where $I^{\rm (c)}_{\rm L2}$ and $I^{\rm (c)}_{\rm R2}$ are charge currents in the drag circuit.
Notice that the charge current is conserved separately on the upper and lower circuit ($I^{\rm (c)}_{\rm L1}+I^{\rm (c)}_{\rm R1}=0$
and $I^{\rm (c)}_{\rm L2}+I^{\rm (c)}_{\rm R2}=0$, respectively).
We will focus on the following two cases:
\begin{itemize}
\item[i)] A pair of capacitively-coupled metallic islands in the Coulomb blockade regime\\
\item[ii)] Two parallel, one-channel, quantum wires.\\
\end{itemize}
 In order to present a more general scenario of the physics of thermal drag in electronic conductors, we choose these two systems since they represent two drastically different physical situations. In particular, while in the quantum wire case the two coupled wires exchange both energy and momentum, in the metallic island case only the former can be transferred. Regarding system i), we study the general properties of the dragged heat using a master equation approach up to second order tunnelling events
(co-tunneling)\cite{Walldorf2017}. We find that the dragged heat current $I^{\rm (h)}_{\rm drag}$ is finite, even in the cases where the dragged charge vanishes
(i.~e. when the island-electrode couplings are energy-independent). We study the behavior of the dragged heat current, driven by either a voltage bias
$V$ or a thermal bias $\Delta T$, as a function of the various parameters characterizing the system, such as the gate voltages and the capacitive
coupling $C_{\rm I}$ between the islands. We find, in particular, that $I^{\rm (h)}_{\rm drag}$ exhibits a maximum as a function of $C_{\rm I}$.
By expanding the dragged heat current for small values of $V$ or $\Delta T$, we find analytic expressions for $I^{\rm (h)}_{\rm drag}$ which result
quadratic in $V$ or $\Delta T$. We find, moreover, that co-tunneling events yield an important impact on the dragged heat current, though not changing
the quadratic dependence on $V$ or $\Delta T$.
Finally, we find that the behavior of the dragged heat current can change qualitatively if one replaces
one of the electrodes in the drag circuit with a superconductor.
More precisely, under appropriate conditions we find that heat can be extracted from the normal electrode in the drag circuit ($I^{\rm (h)}_{\rm L2}<0$).
Additionally, the superconductor allows a finite dragged charge current whose sign can be controlled by the gate voltages.

As far as system ii) is concerned, we carry out an analysis of the system composed of two parallel interacting quantum wires in the low temperature regime,
where the bosonization theory applies. We assume homogeneous intra-wire interaction and inter-wire interaction occurring over a length much
smaller than the total length of the wires. Here we derive an analytic formula, using the bosonization technique in the weak-coupling limit, for the
thermal trans-resistivity $\rho^{\rm (h)}_{12}$, which quantifies the drag of heat (at open circuit) in response to a small temperature difference in the drive circuit.
By comparing $\rho^{\rm (h)}_{12}$ with the electric trans-resistivity $\rho^{\rm (c)}_{12}$, we establish a sort of Wiedemann-Franz law for drag for which
$\rho^{\rm (c)}_{12}/\rho^{\rm (h)}_{12}\propto T^3$, where $T$ is the reference temperature.
This generalizes to drag currents a well known
relation for drive currents \cite{KaneFisher}. Moreover, we identify two characteristic temperature scales: $T_0$, which is associated to the characteristic
wavevector of the coupling, and $T_1$ (with $T_1<T_0$), which is associated to the difference between the Luttinger velocities of the two quantum wires.
We find that the thermal trans-resistivity is linear in $T$ for $T \ll T_1$, decreases like $1/T$ for $T_1\ll T \ll T_0$, and is proportional to $1/T^3$ for $T_0 \ll T$.
Finally, we complement our analysis by performing numerical simulations,
We use a computational protocol that rely on the matrix product states (MPS)
formalism \cite{Sch:wol,Maz:ros,Karrasch} and allows to access the strong coupling and high temperature regime, while validating the weak coupling results.

The paper is organized as follows: In the next Section we will discuss the case i) in which the thermal drag occurs in the case of two coupled metallic islands. We will consider the contribution to the drag due to sequential tunneling and co-tunneling. In Section~\ref{QW} we move to consider the second setup of
two-coupled quantum wires.

\section{Capacitively-coupled metallic islands}
\label{theo}
\begin{figure}[H]
\includegraphics[width=0.7\columnwidth,clip=true]{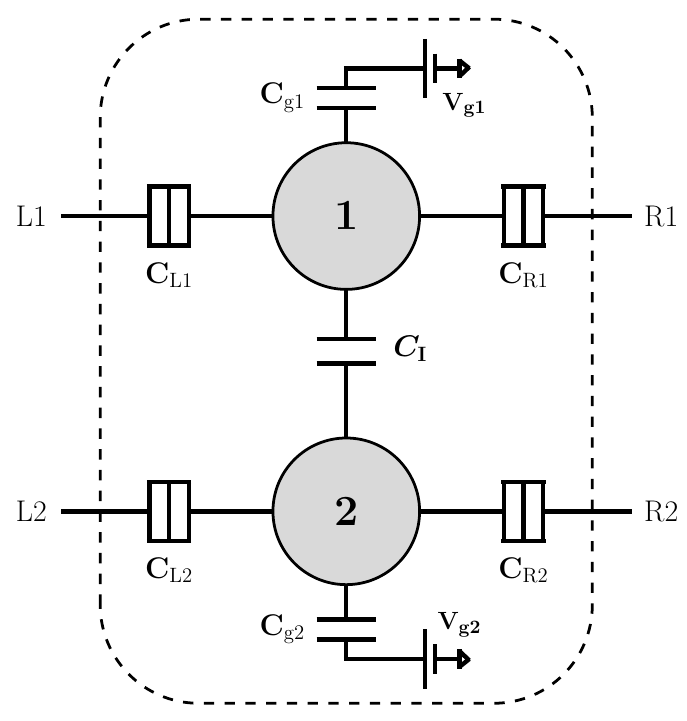}
\centering
\caption{Sketch of the first system under consideration composed of two capacitively-coupled metallic islands labeled by 1 (in the drive circuit) and 2 (in the drag circuit). L1, L2, R1 and R2 labels the four electrodes which are tunnel-coupled to the islands.}
\label{setupSET}
\end{figure}

The first system considered, depicted in Fig.~\ref{setupSET}, consists of two metallic islands (labeled 1 and 2), each one tunnel-coupled to two electrodes and capacitively-coupled to a gate kept at a voltage $V_{{\rm g}i}$, with $i=1,2$.
$C_{\alpha}$ is the capacitance and ${\cal R}_{\alpha}$ is the resistance associated to the tunnel junction between lead $\alpha=$L$i$, R$i$ and the island $i$, while $C_{{\rm g}i}$ is the capacitance associated to the gate.
The two metallic islands (assumed to be at equilibrium temperature $T$) are coupled through a capacitance $C_{\rm I}$, which does not allow electron transfer.
We assume that all capacitances are small so that the charging energies relevant for transport (see below) are the largest energy scales in the system and the islands are in the Coulomb blockade regime.
Single electron tunneling processes in each metallic island, thus, are associated to an increase or decrease in the electrostatic energy of the system, which is given by
\begin{equation}
\begin{aligned}
U(n_1,n_2)=E_{C,1}\left(n_{1}-n_{x_{1}}\right)^{2}+E_{C,2}\left(n_{2}-n_{x_{2}}\right)^{2}+\\
+{E}_{\rm I}\left(n_{1}-n_{x_{1}}\right)\left(n_{2}-n_{x_{2}}\right) .
\end{aligned}
\end{equation}
Here $n_{1}$ and $n_{2}$ represent the number of electrons present on island 1 and 2, respectively.
$E_{C,i}=e^2/(2C_i)$ is the charging energy of island $i$ (where $C_i=C_{{\rm L} i}+C_{{\rm R} i}+C_{{\rm g} i}+C_{{\rm I},i}$, with $C_{\rm I,1}^{-1}=\tilde{C}_2^{-1}+C_{\rm I}^{-1}$, $C_{{\rm I,2}}^{-1}=\tilde{C}_1^{-1}+C_{\rm I}^{-1}$ and $\tilde{C}_i=C_{{\rm L} i}+C_{{\rm R} i}+C_{{\rm g} i}$) and $E_{\rm I}$ is the inter-island interaction energy given by $E_{\rm I}=e^2(\tilde{C_1}+\tilde{C_2}+\tilde{C_1}\tilde{C_2}/C_{\rm I})^{-1}$.
The symbols $n_{x1}$ and $n_{x2}$ represent the ``external charges'' determined by the gate potentials, $V_{\rm g1}$ and $V_{\rm g 2}$ respectively, and dependent on the voltage bias $V$ as
\begin{equation}
n_{x1}=\frac{V/2\; C_{\rm L1}-V/2\; C_{\rm R1}+V_{\rm g1}C_{\rm g1}}{e}
\end{equation}
and
\begin{equation}
n_{x2}=\frac{V_{\rm g2}C_{\rm g2}}{e} .
\end{equation}
For the sake of simplicity, we will assume that $C_{\rm g1}=C_{\rm g2}\equiv C_{\rm g}$ and that all the capacitances relative to the tunnel junctions are equal, namely $C_{\rm L1}=C_{\rm R1}=C_{\rm L2}=C_{\rm R2}$, so that $C_1=C_2\equiv C$ and we can define the charging energy $E_C=e^2/(2C)$.
Note that $n_{x1}$ becomes independent of $V$ and takes the same form as $n_{x2}$.

Charge $I^{\rm (c)}_{\alpha}$ and heat $I^{\rm (h)}_{\alpha}$ currents can be expressed in terms of the probability for the occupation of the islands, and the transition rates for electrons to be exchanged between the islands and an electrodes.
The actual expressions for the currents depend on whether one has to account for only first order tunneling processes (sequential tunneling regime) or second order processes have to be considered too (co-tunneling).
The probability $p(n_1,n_2)$ for the occupation of island 1 with $n_1$ electrons and island 2 with $n_2$ electrons is determined through a set of master equations (see App.~\ref{ME}) which accounts for all possible tunneling processes in the system.

\subsection{Sequential tunneling regime}
\label{seq}
Within the sequential tunneling regime, we have that the charge and heat currents in the lower circuit take the form
\begin{widetext}
\begin{equation}
\label{Ich}
\begin{aligned}
I^{\rm (c/h)}_{\alpha}=Q^{\rm (c/h)}\Big[\Gamma^{\rm (c/h)}_{\alpha,2}(n_1,n_2)\, p(n_1,n_2)+\Gamma^{\rm (c/h)}_{\alpha,2}(n_1+1,n_2)\,  p(n_1+1,n_2)
- \Gamma^{\rm (c/h)}_{2,\alpha}(n_1,n_2)\,  p(n_1,n_2+1) \\
 - \Gamma^{\rm (c/h)}_{2,\alpha}(n_1+1,n_2)\,  p(n_1+1,n_2+1)\Big] ,
\end{aligned}
\end{equation}
\end{widetext}
respectively, where $\alpha={\rm L}2$, ${\rm R}2$ and $Q^{\rm (c)} = e$, $Q^{(h)}=1$.
We have assumed small temperatures and biases so that only four charge states contribute to transport, namely $(n_1,n_2)$, $(n_1+1,n_2)$, $(n_1,n_2+1)$ and $(n_1+1,n_2+1)$.
In Eq.~(\ref{Ich}), $\Gamma^{\rm (c/h)}_{\alpha,i}(n_1,n_2)$ is the particle/heat transition rate for an electron to reach island $i$ from lead $\alpha$ [with the island initially in the state $(n_1,n_2)$], and $\Gamma^{\rm (c/h)}_{i,\alpha}(n_1,n_2)$ is the particle/heat transition rate for an electron leaving island $i$ to reach lead $\alpha$ [with the island in the final state $(n_1,n_2)$].
As long as the energy-dependence of the lead-island couplings~\cite{nota_endep} can be disregarded (see Sec.~\ref{endep}, where this assumption will be lifted),
the particle and heat transition rates can be written as
\begin{equation}
\label{gamma1}
\begin{aligned}
\Gamma^{\rm (c/h)}_{\alpha, i}(n_1,n_2)=\frac{1}{e^2{\cal R}_{\alpha}}F_{\alpha i}^{\rm (c/h)}[\delta U_i(n_1,n_2)-eV_{\alpha}],
\end{aligned}
\end{equation}
and
\begin{equation}
\label{gamma2}
\begin{aligned}
\Gamma^{\rm (c/h)}_{i,\alpha}(n_1,n_2)=\frac{1}{e^2{\cal R}_{\alpha}}G_{i\alpha}^{\rm (c/h)}[\delta U_i(n_1,n_2)-eV_{\alpha}].
\end{aligned}
\end{equation}
In Eqs.~(\ref{gamma1}) and (\ref{gamma2}) the functions $F^{\rm (c/h)}_{\alpha i}$ and $G^{\rm (c/h)}_{i\alpha}$ are defined as
\begin{equation}
\label{fu1}
\begin{aligned}
F^{\rm (c/h)}_{\alpha i}(E)=\int_{-\infty}^{+\infty}    d\epsilon \, z^{\rm (c/h)}f_{\alpha}(\epsilon) [1-f_{i}(\epsilon-E)] ,
\end{aligned}
\end{equation}
\begin{equation}
\label{fu2}
\begin{aligned}
G^{\rm (c/h)}_{i\alpha}(E)=\int_{-\infty}^{+\infty}  d\epsilon \, z^{\rm (c/h)}f_{i}(\epsilon-E) [1-f_{\alpha}(\epsilon)] ,
\end{aligned}
\end{equation}
where $f_{k}(\epsilon)=(1+e^{\epsilon/k_BT_{k}})^{-1}$ is the Fermi distribution at temperature $T_k$, and $z^{\rm (c)}=1$, $z^{\rm (h)}=\epsilon$.
The two quantities
\begin{align}
&\delta U_1(n_1,n_2)=U(n_1+1,n_2)-U(n_1,n_2)\nonumber \\
&\delta U_2(n_1,n_2)=U(n_1,n_2+1)-U(n_1,n_2)
\label{electrocol}
\end{align}
represent the jumps in the electrostatic energy related to the transitions
[note that they appear in Eqs.~(\ref{fu1}-\ref{fu2}) as chemical potentials of the islands].
In the case where all temperatures are equal to $T$, Eqs.~(\ref{gamma1}) and (\ref{gamma2}) for the charge reduce to
\begin{equation}
\Gamma^{\rm (c)}_{\alpha,2}(n_1,n_2)=\frac{1}{e^2{\cal R}_{\alpha}}\frac{\delta U_2(n_1,n_2)}{\exp\left[ {\frac{\delta U_2(n_1,n_2)}{k_BT}}\right]-1},
\end{equation}
and
\begin{equation}
\Gamma^{\rm (c)}_{2,\alpha}(n_1,n_2)=\frac{1}{e^2{\cal R}_{\alpha}}\frac{-\delta U_2(n_1,n_2)}{\exp\left[ {\frac{-\delta U_2(n_1,n_2)}{k_BT}}\right]-1}.
\end{equation}

The assumption of energy-independent couplings allows us to make general statements thanks to the fact that the currents $I^{\rm (c/h)}_{\alpha}$ are proportional to $1/{\cal R}_{\alpha}$.
In the lower circuit, in particular, the proportionality constants are equal for the two leads (i. e., $I^{\rm (c)}_{\rm L2}{\cal R}_{\rm L2}=I^{\rm (c)}_{\rm R2}{\cal R}_{\rm R2}$ and $I^{\rm (h)}_{\rm L2}{\cal R}_{\rm L2}=I^{\rm (h)}_{\rm R2}{\cal R}_{\rm R2}$) since no biases are applied.
As far as charge is concerned, current conservation in the lower circuit ($I^{\rm (c)}_{\rm L2}+I^{\rm (c)}_{\rm R2}=0$) implies that the individual charge currents in the lower circuit vanish identically, and therefore $I^{\rm (c)}_{\rm drag}$ is zero even in the case of asymmetric barriers  (${\cal R}_{\rm L2}\ne {\cal R}_{\rm R2}$).
On the other hand, no conservation holds for the heat currents~\cite{heat_con} in the lower circuit so that the two heat currents, $I^{\rm (h)}_{\rm L2}$ and $I^{\rm (h)}_{\rm R2}$, are in general non-vanishing.
In particular, for symmetry reasons they are equal when ${\cal R}_{\rm L2}= {\cal R}_{\rm R2}$, and therefore $I^{\rm (h)}_{\rm drag}$ is finite only in the case of asymmetric barriers.
The presence of heat currents in the lower circuit is a result of the energy transferred from the upper circuit, thanks to the capacitive coupling.
Indeed, as detailed in the following, this energy transfer occurs through the dependence of $\delta U_2$, which controls the transition rates for the lower island, on the charge state of the upper island $n_1$, see Eq.~(\ref{electrocol}).

\begin{figure}[H]
\includegraphics[width=\columnwidth,clip=true]{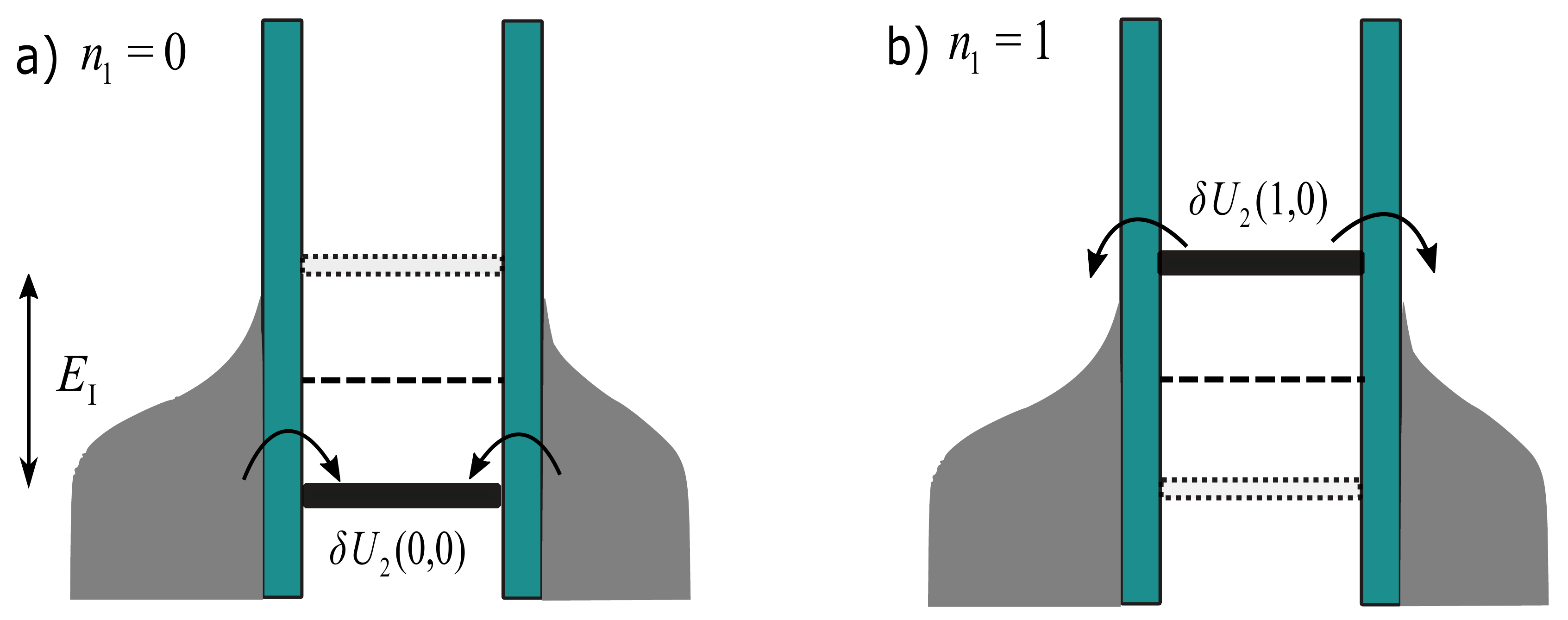}
\caption{Energies scheme for the lower, drag, circuit. Green rectangles represent the tunnel barriers. Grey areas represent the Fermi distribution functions of the leads, whose common equilibrium electrochemical potential, set to zero, is indicated by a dashed line. Thick horizontal black lines indicate the position of the chemical potential of the island for $n_{x1}=n_{x2}=1/2$. a) The upper island is empty and the chemical potential is $\delta U_2(0,0)$: electrons can jump on the island. b) The upper island is occupied and the chemical potential is $\delta U_2(1,0)$: electrons can jump out of the island.}
\label{energies1}
\end{figure}

For the sake of definiteness, let us assume that the relevant charge states are $(0,0)$, $(0,1)$, $(1,0)$ and $(1,1)$.
Thus the jumps in electrostatic energy related to the currents in the lower island are
\begin{equation}
\label{u200}
\delta U_2(0,0)=E_C(1-2n_{x2})-E_{\rm I}n_{x1} ,
\end{equation}
for the case where the upper island is empty, and
\begin{equation}
\label{u210}
\delta U_2(1,0)=E_C(1-2n_{x2})+E_{\rm I}(1-n_{x1}) ,
\end{equation}
for the case where the upper island is occupied.
Equations~(\ref{u200}) and (\ref{u210}) express the fact that the position of the two chemical potentials $\delta U_2(0,0)$ and $\delta U_2(1,0)$ of the lower island, with respect to common equilibrium electrochemical potential of the lower leads (set to zero), is expressed in terms of $n_{x1}$ and $n_{x2}$.
When $n_{x1}=n_{x2}=1/2$ we obtain $\delta U_2(0,0)=-E_{\rm I}/2$ and $\delta U_2(1,0)=+E_{\rm I}/2$.
The energy scheme for the lower island is represented in Fig.~\ref{energies1}a) for the former case and in panel Fig.~\ref{energies1}b) for the latter case.
If we assume a small temperature $T$, an electron can jump on the island from one of the electrodes only when the upper island is empty, since the corresponding chemical potential $\delta U_2(0,0)$ is below the electrochemical potential of the leads, see panel a).
Such an electron can jump out of the island only when the upper island gets occupied, since the chemical potential $\delta U_2(1,0)$ is now greater than zero.
This sequence of processes allows the heat currents $I^{\rm (h)}_{L2}$ and $I^{\rm (h)}_{R2}$ to be finite as long as the interaction energy $E_{\rm I}\ne 0$.
Such heat currents can be modulated by varying $n_{x1}$ and $n_{x2}$, which produces a rigid shift of the position of the two chemical potentials $\delta U_2(0,0)$ and $\delta U_2(1,0)$, see Eqs.~(\ref{u200}) and (\ref{u210}).
Note that the difference $\delta U_2(1,0) - \delta U_2(0,0)=E_{\rm I}$, independently of $n_{x1}$ and $n_{x2}$.

Analytic, even though cumbersome, expressions for the heat currents could be derived in the limit of small biases $V$ and $\Delta T$.
Interestingly, heat currents turn out to be second order in $V$ and $\Delta T$ (note that the heat currents in the upper circuit are first order in $V$ and $\Delta T$).
In particular, when $n_{x1}=n_{x2}=1/2$, the dragged heat current takes the simple form
\begin{equation}
\label{heatV}
I^{\rm (h)}_{\rm drag}=\frac{\xi {\cal R}_\parallel}{16{\cal R}}\left[\frac{1}{{\cal R}_{\rm L2}}-\frac{1}{{\cal R}_{\rm R2}}\right] \csch\xi \left[\xi{\csch}\xi-\sech\xi\right]{ V}^{2} ,
\end{equation}
when $\Delta T = 0$ and expanding in $V/E_C$, while
\begin{widetext}
\begin{equation}
\label{heatT}
I^{\rm (h)}_{\rm drag}=\frac{\xi {\cal R}_\parallel}{6e^{2}{\cal R}}\left[\frac{1}{{\cal R}_{\rm L2}}-\frac{1}{{\cal R}_{\rm R2}}\right] \csch\xi  \left[2\xi\left(\frac{\pi^{2}}{4}+\xi^{2}\right)\csch\xi-\left(\frac{\pi^{2}}{2}+3\xi^{2}\right)\sech\xi\right](k_B\Delta T/2)^{2} ,
\end{equation}
\end{widetext}
when $V=0$ and expanding in $\Delta T/T$ (only the leading terms in $k_B\Delta T/E_C$ are retained)\cite{footnote}.
In Eqs.~(\ref{heatV}) and (\ref{heatT}) we have defined $\xi=E_{\rm I}/(4k_BT)$ and
\begin{equation}
{\cal R}_\parallel =\left(\frac{2}{{\cal R}}+\frac{1}{{\cal R}_{\rm L2}}+\frac{1}{{\cal R}_{\rm R2}}\right)^{-1} ,
\end{equation}
and assumed ${\cal R}_{\rm L1}={\cal R}_{\rm R1}={\cal R}$.
Eqs.~(\ref{heatV}) and (\ref{heatT}) show that the dragged heat current is finite only when the interaction energy $E_{\rm I}\ne 0$ and depends on $E_{\rm I}$ only through the ratio $E_{\rm I}/(k_BT)$ (this is true only when $n_{x1}=n_{x2}=1/2$).
Moreover, we mention that in the presence of both voltage and thermal biases the contribution to the dragged heat current is proportional to the product $V \Delta T$ and exhibits the same qualitative behavior as for the voltage or thermal bias only case. For $n_{x1} = n_{x2} = 1/2$ such contribution vanishes.
\begin{figure}[ht]
\includegraphics[width=\columnwidth,clip=true]{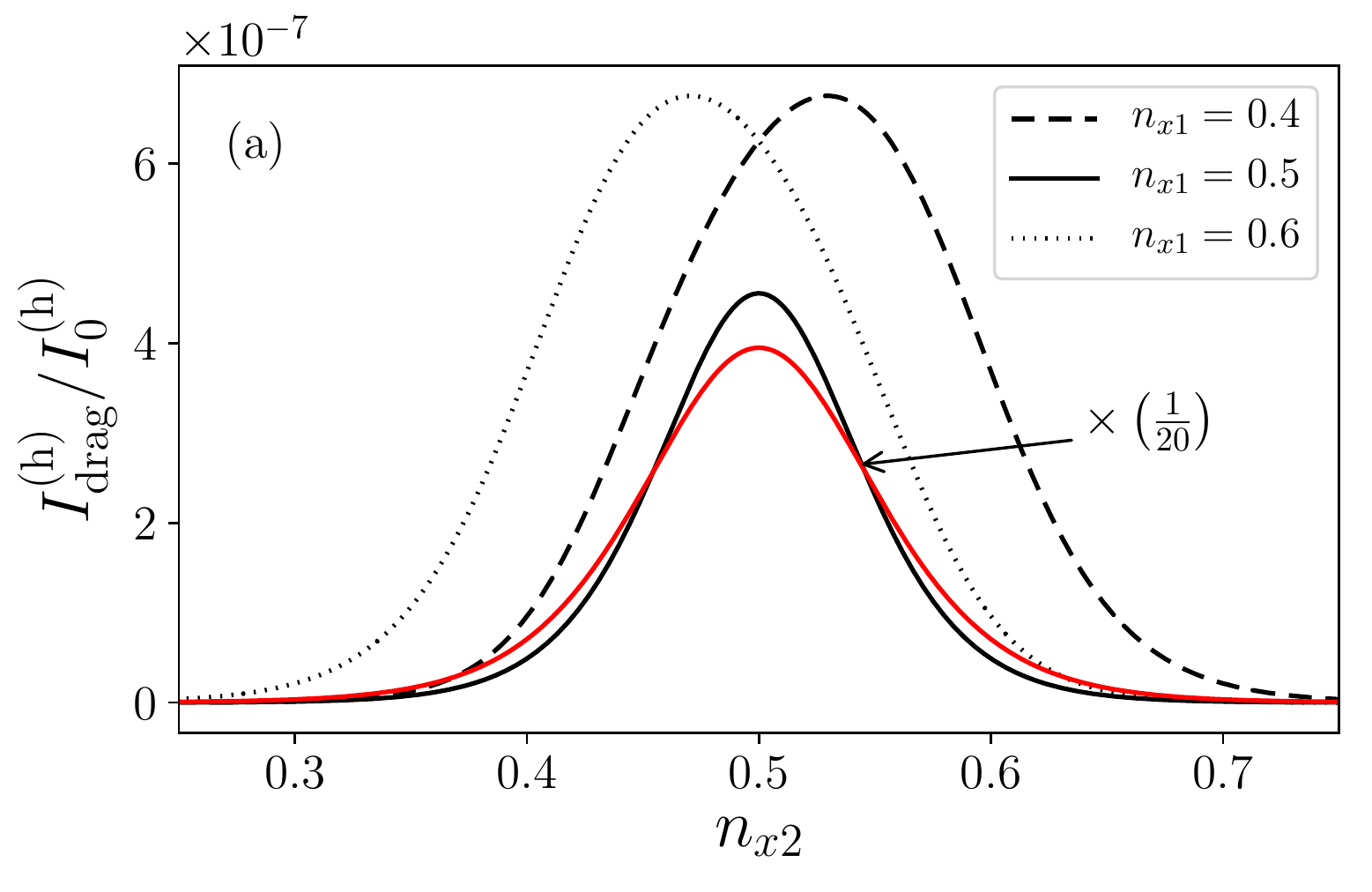}
\includegraphics[width=\columnwidth,clip=true]{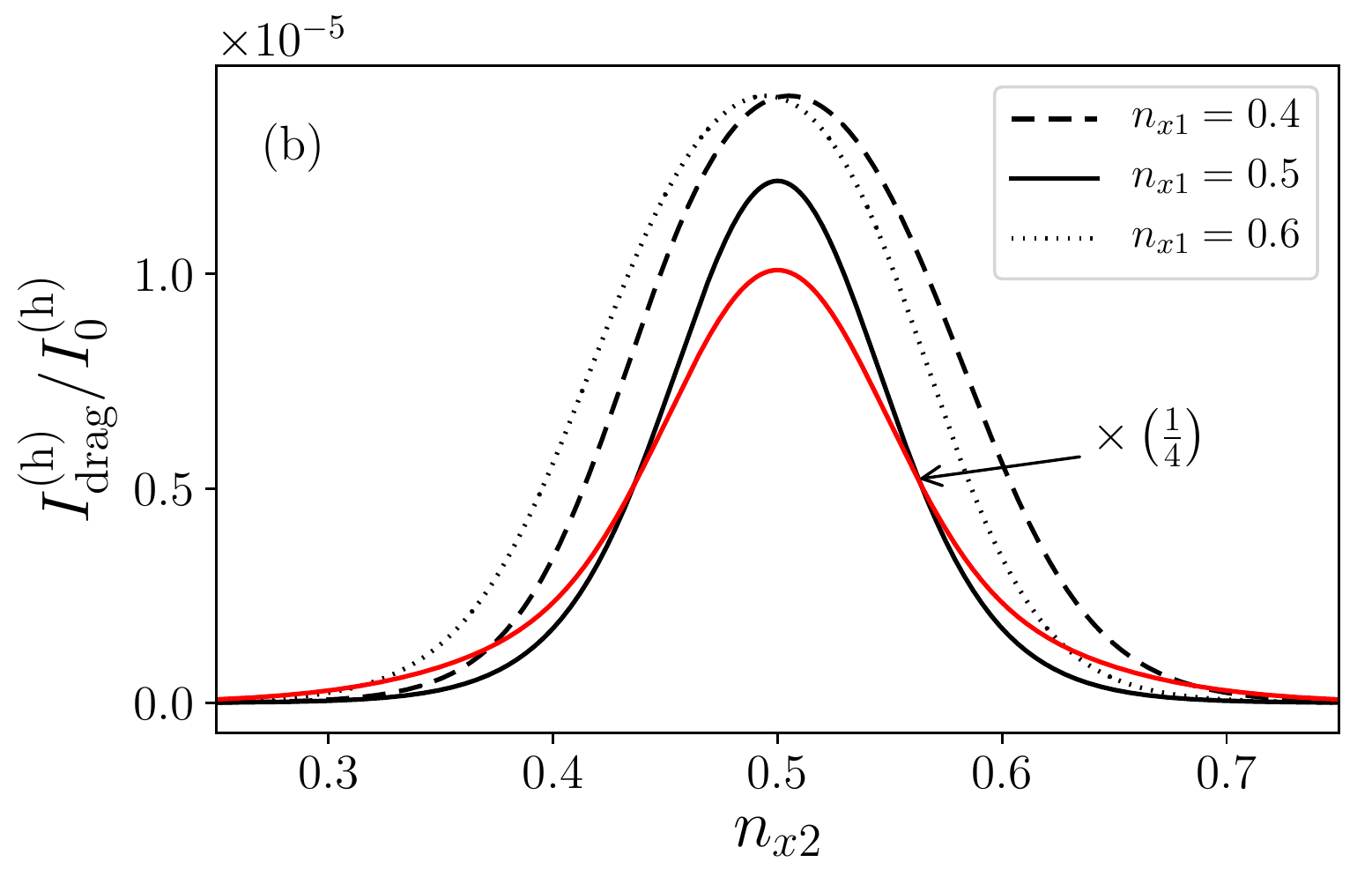}
\caption{Dragged heat currents plotted as functions of $n_{x2}$ for different values of $n_{x1}$. The results accounting for sequential tunneling only are plotted in black, while results including co-tunneling contributions are plotted in red.
${\cal R}_{\rm L1}= {\cal R}_{\rm R1}={\cal R}_{\rm L2}=5 {\cal R}_{\rm Q}$, ${\cal R}_{\rm R2}=10 {\cal R}_{\rm Q}$, $k_{\rm B}T=0.05 E_C$, and $E_{\rm I}=0.4 E_C$.
(a) $\Delta T=0$ and $V=0.08 E_C/e$; (b) $V=0$ and $\Delta T=0.08 E_C/k_B$.
The solid curves have been multiplied by a factor $1/20$ in panel (a) and by a factor $1/4$ in panel (b).
The heat current is given in units of $I^{\rm (h)}_0=e^2/(4C^2{\cal R})$.}
\label{fig.Ihvsnx2}
\end{figure}
More manageable expressions can be obtained by further expanding in powers of the interaction energy $E_{\rm I}$, namely we get
\begin{equation}
I_{\rm drag}^{\rm (h)}= \left[\frac{1}{{\cal R}_{\rm L2}}-\frac{1}{{\cal R}_{\rm R2}}\right]\frac{{\cal R}_\parallel\xi^{2}}{48{\cal R}}V^{2} ,
\end{equation}
when $\Delta T=0$, and
\begin{equation}
I_{\rm drag}^{\rm (h)}=\left[\frac{1}{{\cal R}_{\rm L2}}-\frac{1}{{\cal R}_{\rm R2}}\right]\frac{\left(\pi^{2}-6\right){\cal R}_\parallel\xi^{2}}{36e^{2}{\cal R}}\left(k_B\Delta T/2\right)^{2}
\end{equation}
when $V=0$.
Note that both expressions are second order in $E_{\rm I}$.

Let us now concentrate on the dependence of the dragged heat current $I^{\rm (h)}_{\rm drag}$ on the external charges, i.~e. on the gate voltages, and show numerical results for the asymmetric barriers case specified by ${\cal R}_{\rm R2}=10 {\cal R}_{\rm Q}$, ${\cal R}_{\rm L2}=5 {\cal R}_{\rm Q}$ (${\cal R}_{\rm Q}=e^2/h$ is the resistance quantum), while setting $k_BT=0.05 E_C$, ${\cal R}={\cal R}_{\rm L1}= {\cal R}_{\rm R1}=10 {\cal R}_{\rm Q}$, and $E_{\rm I}=0.4 E_C$.
In Fig.~\ref{fig.Ihvsnx2} we plot the dragged heat current as a function of $n_{x2}$ (determined by the gate voltage acting on island 2) for three different fixed values of $n_{x1}$.
The black curves accounts for sequential tunneling processes only (solid $n_{x1}=0.5$, dashed $n_{x1}=0.4$ and dotted $n_{x1}=0.6$), while the red solid curve accounts also for co-tunneling contributions (see below).
In Fig.~\ref{fig.Ihvsnx2}a) and \ref{fig.Ihvsnx2}b) the currents are, respectively, a result of a voltage bias $V$ (with $\Delta T=0$) or a thermal bias $\Delta T$ (with $V=0$).

Fig.~\ref{fig.Ihvsnx2}a) shows that when $n_{x1}=1/2$, solid curve, $I^{\rm (h)}_{\rm drag}$ exhibits a peak at $n_{x2}=1/2$, while the peak is shifted to a larger (smaller) value of $n_{x2}$ when $n_{x1}=0.4$ ($n_{x1}=0.6$).
This effect can be understood by noticing that the dragged heat current is expected to be maximal when the two chemical potentials of the lower island $\delta U_2(1,0)$ and $\delta U_2(0,0)$ are {\it equidistant} with respect to the equilibrium electrochemical potential set by the electrodes [see Fig.~\ref{energies1}].
In this case, in fact, the heat transition rate for an electron to enter the island from the left lead (non-vanishing only if $n_1=0$) is equal to the heat transition rate for an electron to leave the island to go to the left lead (non-vanishing only if $n_1=1$).
By departing from the equidistant configuration, one of the two rates gets suppressed resulting in a suppression of the heat current\cite{ratesnx2}.
For $n_{x1}=1/2$ the equidistant configuration occurs when $n_{x2}=1/2$, while when $n_{x1}=0.4$ ($n_{x1}=0.6$) the equidistant configuration occurs when $n_{x2}=0.5 + 0.05\, E_{\rm I}/E_{\rm C}>1/2$ ($n_{x2}=0.5 - 0.05\, E_{\rm I}/E_{\rm C}<1/2$).
Notice that the value of $I^{\rm (h)}_{\rm drag}$ is over one order of magnitude bigger in the case $n_{x1}=1/2$, with respect to the cases $n_{x1}=0.4$ and $n_{x1}=0.6$.
The reason for this behavior is related to the fact that in the former case the heat current in the drive circuit (and therefore the energy transferred in the lower circuit) is maximum.

We checked that the position and the shape of the peaks does not change by varying the value of $V$, while the maximum value increases with it.
On the contrary, an increase in the temperature $T$ produces a proportional increase in the width of the peaks ($\Delta n_{x2}\simeq 2k_BT/E_C$), on the one hand, and a decrease in the separation between the peaks at $n_{x2}=0.4$ and at $n_{x2}=0.6$, on the other.
Thus, temperature seems to have a less intuitive effect on the dragged heat current.
Remarkably, the width of the peaks is virtually independent of $E_{\rm I}$.
In Fig.~\ref{fig.Ihvsnx2}b) we show plots of the dragged heat current in the presence of a thermal bias in the drive circuit.
The behavior of $I^{\rm (h)}_{\rm drag}$ in this case is similar to the one in the presence of a voltage bias, with the following little differences:
i) the value of the heat current for the cases $n_{x1}=0.4$ and $n_{x1}=0.6$ is not dramatically suppressed with respect to the $n_{x1}=1/2$ case (a factor 4 with respect to a factor 20); ii) the shift in the positions of the peaks for the cases $n_{x1}=0.4$ and $n_{x1}=0.6$ is smaller with respect to the voltage-bias case.

\begin{figure}[!ht]
\includegraphics[width=\columnwidth,clip=true]{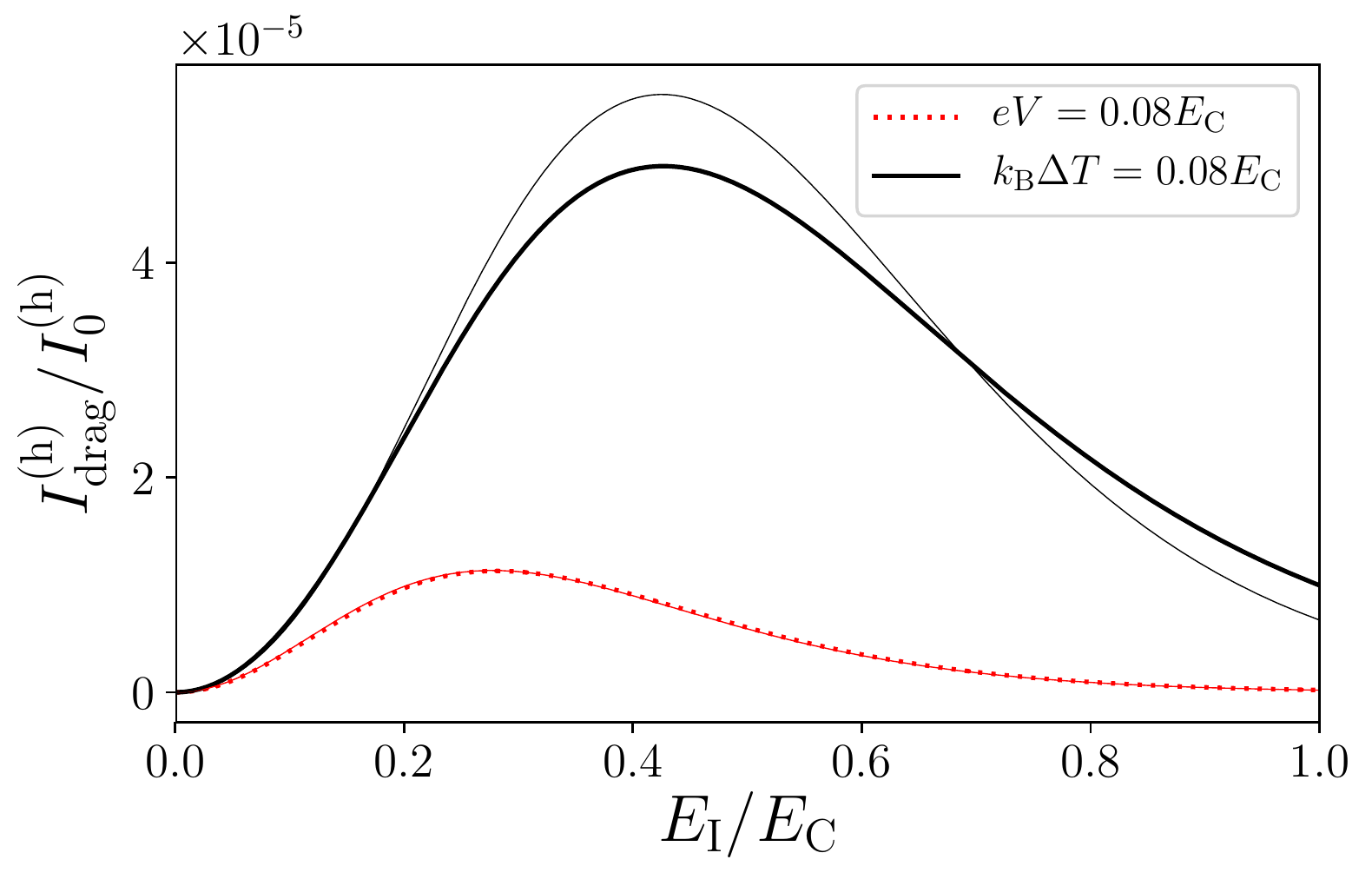}
\caption{Dragged heat current (sequential tunneling only) plotted as a function of $E_{\rm I}$ for the case $V=0.08E_C$ and $\Delta T=0$ (red lines) the case $\Delta T=0.08E_C$ and $V=0$ (black lines) for $n_{x1}=n_{x2}=1/2$. Thin black and red curves are plots of the analytic expressions Eqs.~(\ref{heatV}) and  (\ref{heatT}), respectively.
The other parameters are chosen as follows: ${\cal R}_{\rm L1}= {\cal R}_{\rm R1}={\cal R}_{\rm L2}= 5 {\cal R}_{\rm Q}$, ${\cal R}_{\rm R2}=10 {\cal R}_{\rm Q}$, and $k_{\rm B}T=0.05 E_C$.}
\label{fig.IhvsEI}
\end{figure}

Let us now concentrate on the role of $E_{\rm I}$ on the dragged heat current.
Notice that the interaction energy can be expressed as
\begin{equation}
{E}_{\rm I}=E_C\; \frac{2}{1+\frac{\tilde{C}}{C_{\rm I}}} ,
\end{equation}
where $\tilde{C}\equiv \tilde{C}_1=\tilde{C}_2$, and that it is bounded by the inequality ${E}_{\rm I}\leq 2E_C$.
In Fig.~\ref{fig.IhvsEI}, $I^{\rm (h)}_{\rm drag}$ is plotted as a function of $E_{\rm I}$ for the voltage bias case (dashed red line) and for the thermal bias case (solid black line) for $n_{x1}=n_{x2}=1/2$.
As a general feature, we note that the dragged heat current is maximal for intermediate values of $E_{\rm I}$.
This agrees with the fact that, on the one hand, $I^{\rm (h)}_{\rm drag}$ must decrease for large values of $E_{\rm I}$ as a consequence of the fact that the probability $p(1,1)$, thus the occurrence of the process depicted in Fig.~\ref{energies1}(b), gets suppressed (indeed, $E_{\rm I}$ represents the inter-island Coulomb repulsion which hinders the occupation of the lower island when the upper island is occupied). On the other hand, $I^{\rm (h)}_{\rm drag}$ vanishes for $E_{\rm I}=0$ due to the absence of electrostatic coupling.
In Fig.~\ref{fig.IhvsEI}, while the thick lines are numerical results, the thin lines are the analytical solutions for small voltage and temperature biases, Eqs.~(\ref{heatV}) and (\ref{heatT}).
It worthwhile stressing that while the red curves coincide, the black curves closely match only for $E_{\rm I}<0.2E_C$ and thereafter depart significantly.
This is due to the fact that, despite $k_B\Delta T$ is small with respect to $E_C$, $k_B\Delta T$ is larger than $k_BT$ and Eq.~(\ref{heatT}) does not hold.
Nevertheless, the position of the maxima $E^{\rm max}_{\rm I}$ are well predicted by the analytical expressions, Eqs.~(\ref{heatV}) and (\ref{heatT}), even for larger values of $\Delta T$ and $V$.
The solution of a transcendent equation yield $E^{\rm max}_{\rm I}\simeq 5.5 k_BT$, for the voltage bias case, and $E^{\rm max}_{\rm I}\simeq 8.5 k_BT$, for thermal bias case.
Finally, unlike Eqs.~(\ref{heatV}) and (\ref{heatT}), we notice that $I^{\rm (h)}_{\rm drag}$ at $n_{x1}=n_{x2}=1/2$ for large enough $V$ and $\Delta T$ depends on $E_{\rm I}$ not only through the ratio $E_{\rm I}/(k_BT)$.

\begin{figure}[ht]
\includegraphics[width=\columnwidth,clip=true]{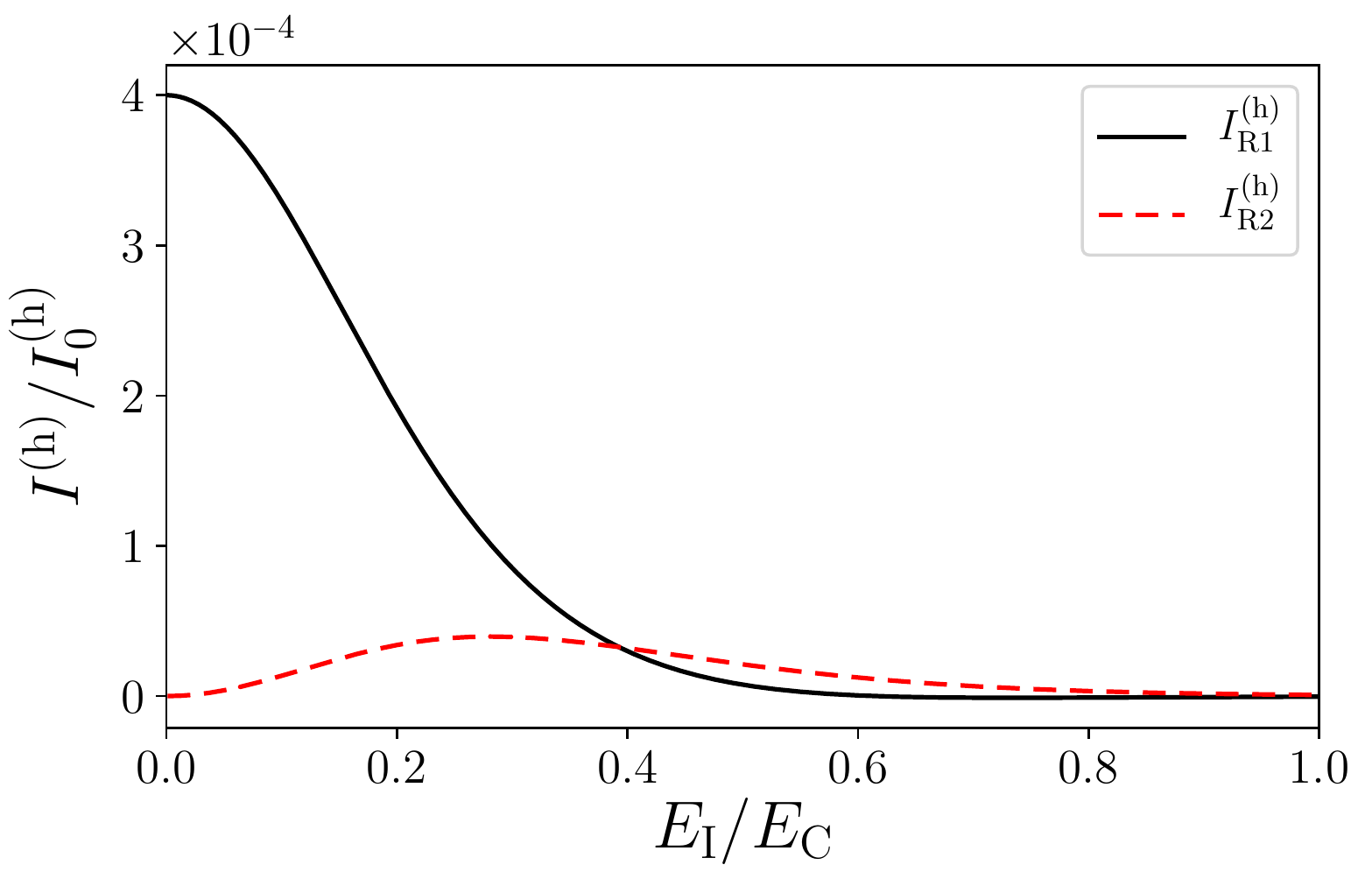}
\caption{Heat currents $I^{\rm (h)}_{\rm R1}$  (solid black curve) and $I^{\rm (h)}_{\rm R2}$ (dashed red curve) plotted as a function of $E_{\rm I}$ for the case $\Delta T=0$ and accounting for sequential tunneling only. The other parameters are chosen as follows: ${\cal R}_{\rm L1}={\cal R}_{\rm L2}= {\cal R}_{\rm R1}={\cal R}_{\rm R2}= 10{\cal R}_{\rm Q}$, $n_{x1}=n_{x2}=1/2$, $V=0.08 E_C/e$, and $k_{\rm B}T=0.05 E_C$.}
\label{fig.IhvsEI2}
\end{figure}

We conclude this section by comparing the heat current in the drive circuit, for example $I^{\rm (h)}_{\rm R1}$, with the one in the drag circuit, for example $I^{\rm (h)}_{\rm R2}$.
In the case of a thermal bias, it turns out that $I^{\rm (h)}_{\rm R2}<I^{\rm (h)}_{\rm R1}$, as expected from the fact that $I^{\rm (h)}_{\rm R1}$ is linear in $\Delta T$, while $I^{\rm (h)}_{\rm R2}$ is quadratic in $\Delta T$ (at least for small values of $\Delta T$).
In the voltage-bias case, surprisingly, we find that $I^{\rm (h)}_{\rm R2}$ is larger than $I^{\rm (h)}_{\rm R1}$ for large enough interaction energy, as shown in Fig.~\ref{fig.IhvsEI2}, where the crossing occurs at $E^{\rm cross}_{\rm I}\simeq 0.4 E_C$.
More precisely, the value of $E^{\rm cross}_{\rm I}$ decreases linearly by decreasing $T$, thereafter saturating, for small $T$, to a finite value of $E^{\rm cross}_{\rm I}\sim eV$, i.~e. very close to the applied voltage.

\subsection{Co-tunnelling contributions}
\label{cot}
When the barriers' resistances do not largely exceed the resistance quantum ${\cal R}_{Q}$ and temperatures are low, it is important to account for second-order tunneling events (co-tunneling contributions).
Given the large number of electrons in the islands, we will only consider inelastic co-tunneling.
Co-tunneling contributions affect the expressions of the currents (\ref{Ich}), see App.~\ref{Curr}, and the master equations, see App.~\ref{ME}, by introducing additional terms.
These are related to the co-tunneling particle and heat transition rates involving an electron entering or leaving island 1 through the upper leads and a second electron entering or leaving island 2 through the lower leads (see App.~\ref{Rate}).
In the present situation, where there are no voltage and temperature biases applied to the drag (lower) circuit, the number of processes that contributes to the current in L$2$ (R$2$) is limited to the ones that involve a tunneling event between island 2 and its lead L$2$ (R$2$) and all possible tunneling events between island 1 and its leads L$1$ and R$1$ (see App.~\ref{Curr} for the expression of the current $I^{\rm (h)}_{\rm R2}$).
We were able to obtain analytical expressions for charge and heat currents only in the voltage-biased case.

Due to the energy-independence of lead-island couplings, also in the presence of co-tunneling contributions the currents $I^{\rm (c/h)}_{\alpha}$ remain proportional to $1/{\cal R}_{\alpha}$ in such a way that the charge currents in the drag circuit vanish also in the case of asymmetric barriers (${\cal R}_{\rm L2}\ne {\cal R}_{\rm R2}$).
On the contrary, the dragged heat currents, which are non-zero even for sequential tunneling, can give rise to quantitatively important changes (when resistances are small and temperatures are low).
As shown in Fig.~\ref{fig.Ihvsnx2} (red curves), co-tunneling gives rise to a broadening and lowering of the peaks with respect to the sequential tunneling only case (solid black curves), both in the voltage and thermal bias cases.

\begin{figure}[ht]
\includegraphics[width=\columnwidth,clip=true]{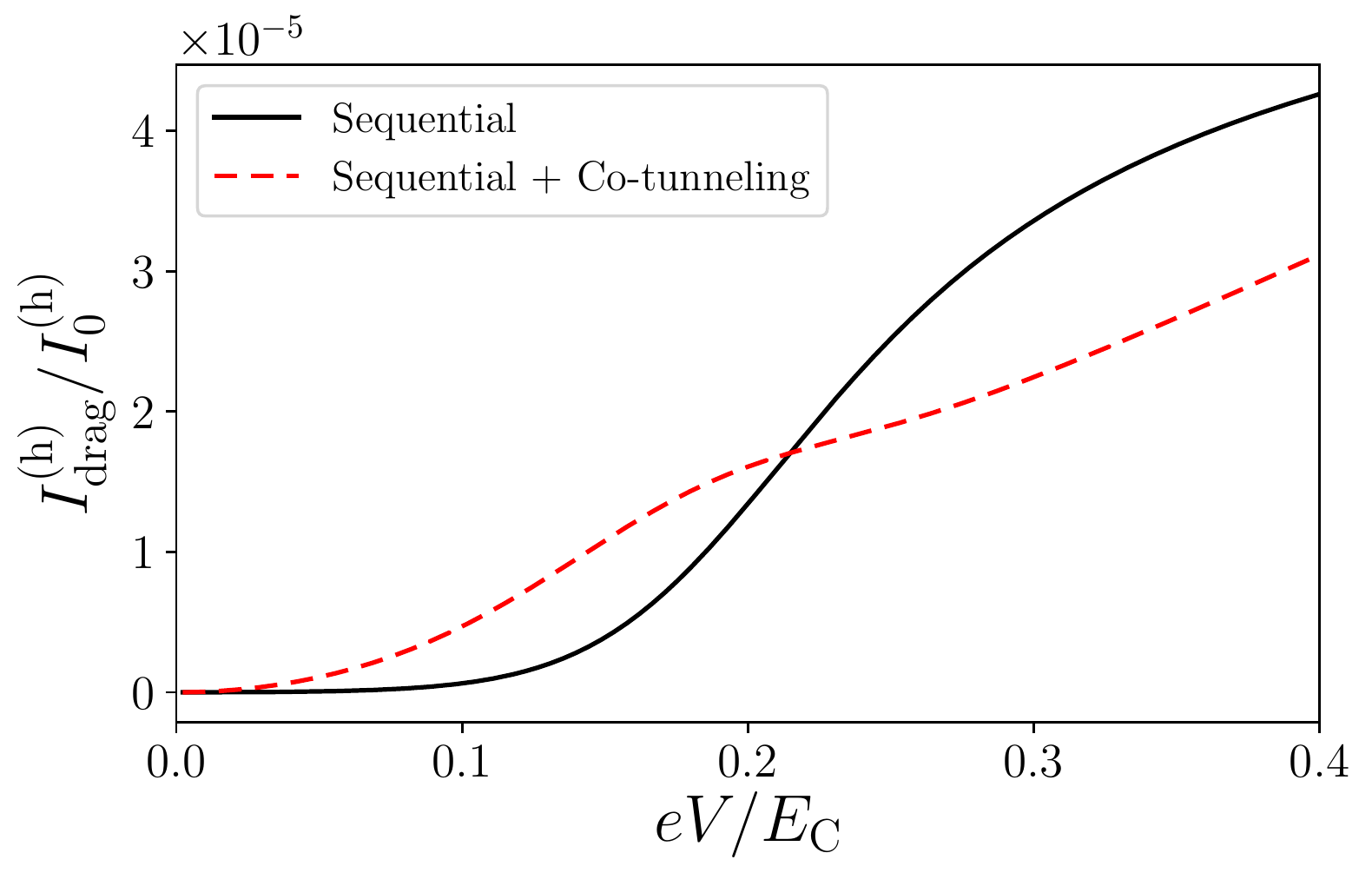}
\caption{Dragged heat current $I^{\rm (h)}_{\rm drag}$ plotted as a function of $V$ accounting for sequential tunneling only (solid black line) and including co-tunneling contributions (dashed red line) for $\Delta T=0$ and $n_{x1}=n_{x2}=0.478$. The other parameters are chosen as follows: ${\cal R}_{\rm L1}= {\cal R}_{\rm R1}={\cal R}_{\rm L2}= 2{\cal R}_{\rm Q}$, ${\cal R}_{\rm R2}=4 {\cal R}_{\rm Q}$, $E_{\rm I}=0.1E_C$ and $k_{\rm B}T=0.01 E_C$.
}
\label{fig.IhvsV}
\end{figure}

In Fig.~\ref{fig.IhvsV}, $I^{\rm (h)}_{\rm drag}$ is plotted as a function of $V$, for $\Delta T=0$, in a wide range of voltages up to $0.4E_C$.
The solid black curve accounts for sequential tunneling only, while the red dashed curve includes co-tunneling events.
Fig.~\ref{fig.IhvsV} shows that co-tunneling events produce an increase of the dragged heat current for values of $V$ in the lower range and a decrease in the upper range. This reflects the fact that, for low voltages, co-tunneling contributions becomes dominant since the Coulomb gap does not allow for first order transport processes (sequential tunneling)~\cite{Keller2016,RSanchez2010}.
Finally, we numerically check that, for small voltage and temperature biases, $I^{\rm (h)}_{\rm drag}$ remains quadratic in $V$ and $\Delta T$ even when co-tunneling contributions are important.
For the specific choice of parameters used in Fig.~\ref{fig.IhvsV}, we get that $I^{\rm (h)}_{\rm drag}$ remains proportional to $V^2$ up to $V\sim 0.05 E_C/e$, when sequential tunneling only is accounted for, while up to $V\sim 0.1 E_C/e$, when co-tunneling is also included.

\subsection{Superconducting electrode}
\label{endep}
In this section we assume that one of the electrodes in the drag circuit (the right-hand one, R2, for definiteness) is superconducting.
This case is interesting since the transition rates cannot be written as in Eqs.~(\ref{gamma1}) and (\ref{gamma2}).
Indeed, the particle transition rates $\Gamma^{\rm (c)}_{R2,2}(n_1,n_2)$ can be written as~\cite{Schon1997}
\begin{align}
\label{gamma1E}
\Gamma^{\rm (c)}_{R2,2}&(n_1,n_2)=\frac{4\pi}{\hbar}\int d\epsilon \int d\epsilon '
\left| t(\epsilon)\right|^{2}\rho_{R2}(\epsilon)\rho_{2}(\epsilon ')f_{R2}(\epsilon)\nonumber \\
&\times \left[1-f_{2}(\epsilon')\right]\delta\left[\epsilon'-\epsilon+\delta U_2(n_1,n_2) \right] ,
\end{align}
where $\rho_{R2}(\epsilon)$ [$\rho_{2}(\epsilon)$] is the density of states (DOS) of the superconducting lead R2 (island $2$) and $t(\epsilon)$ is the tunneling matrix element of the junction (the heat transition rates are defined analogously).
The DOS of the electrode R2 is given by
\begin{equation}
\label{dosSC}
\rho_{{\rm R2}}(\epsilon)=\rho^{\rm nor}_{{\rm R2}} \Theta \left[|\epsilon | - \Delta\right] \frac{|\epsilon |}{\left(\epsilon^{2}-\Delta^{2}\right)^{1/2}} ,
\end{equation}
where $\rho^{\rm nor}_{{\rm R{2}}}$ is the DOS of the electrode in the normal state, $\Theta$ is the Heaviside step function, and $\Delta$ is the  superconducting gap.
We assume that $\Delta\ll E_{C}$, so that Andreev reflection is largely suppressed~\cite{Schon1997}, and ${\cal R}^{\rm nor}_{\rm R2}\gg {\cal R}_{Q}$, so that sequential tunneling of quasi-particles becomes the dominant process.
${\cal R}^{\rm nor}_{\rm R2}$ is defined as the normal state tunnel resistance of the junction R$2$, namely
\begin{equation}
\frac{1}{{\cal R}^{\rm nor}_{R2}}=\frac{4\pi e^{2}}{\hbar}\rho_{2}\,\rho_{R2}^{\rm nor}\left|t\right|^{2} ,
\label{tunnel_res}
\end{equation}
since the tunneling matrix element is energy-independent.
Notice that Eq.~(\ref{gamma1E}) reduces to Eq.~(\ref{gamma1}) when the energy-dependence of the tunneling matrix elements and of the DOS of lead and island can be disregarded.
\begin{figure}[ht]
\includegraphics[width=\columnwidth,clip=true]{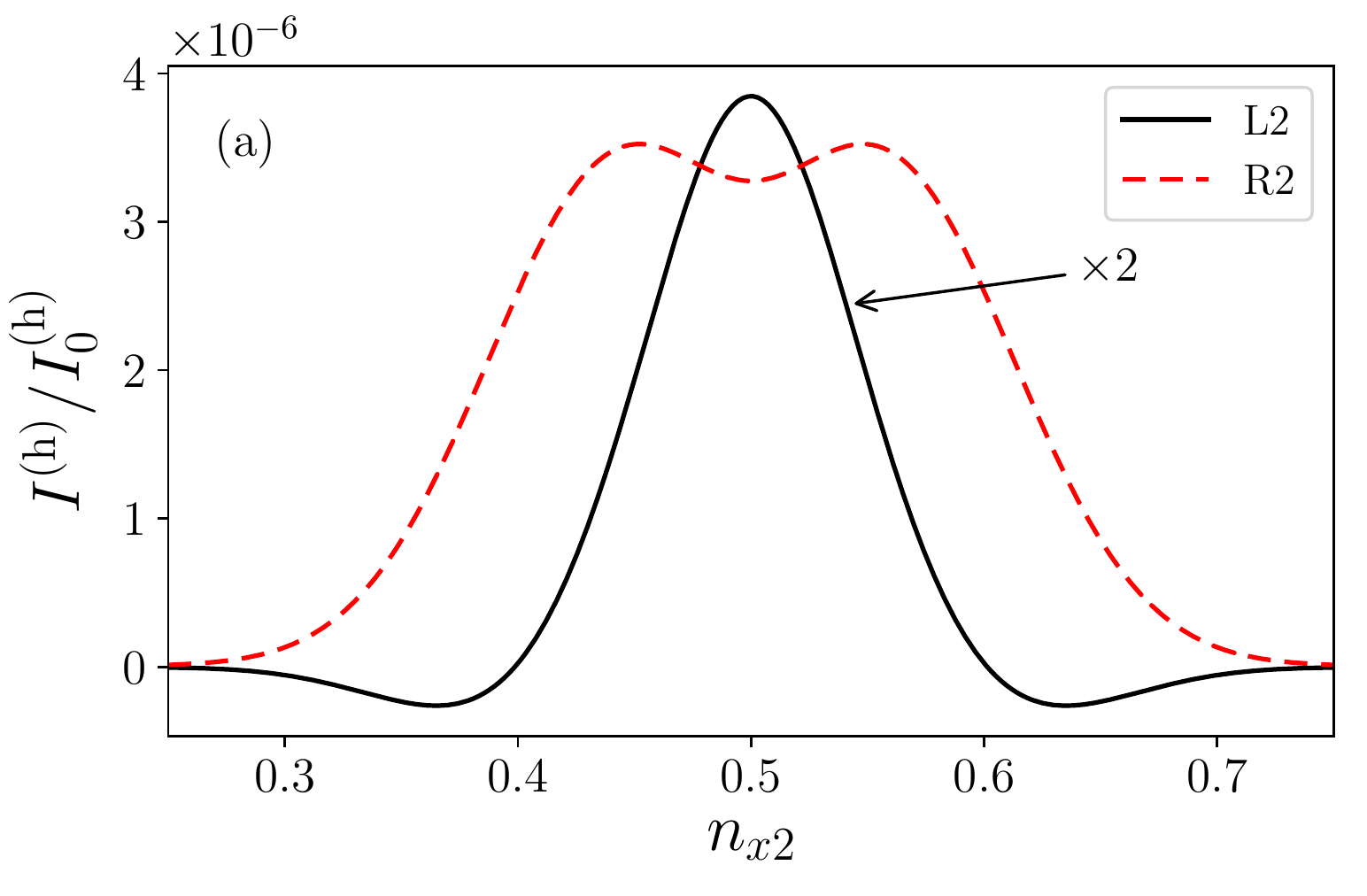}
\includegraphics[width=\columnwidth,clip=true]{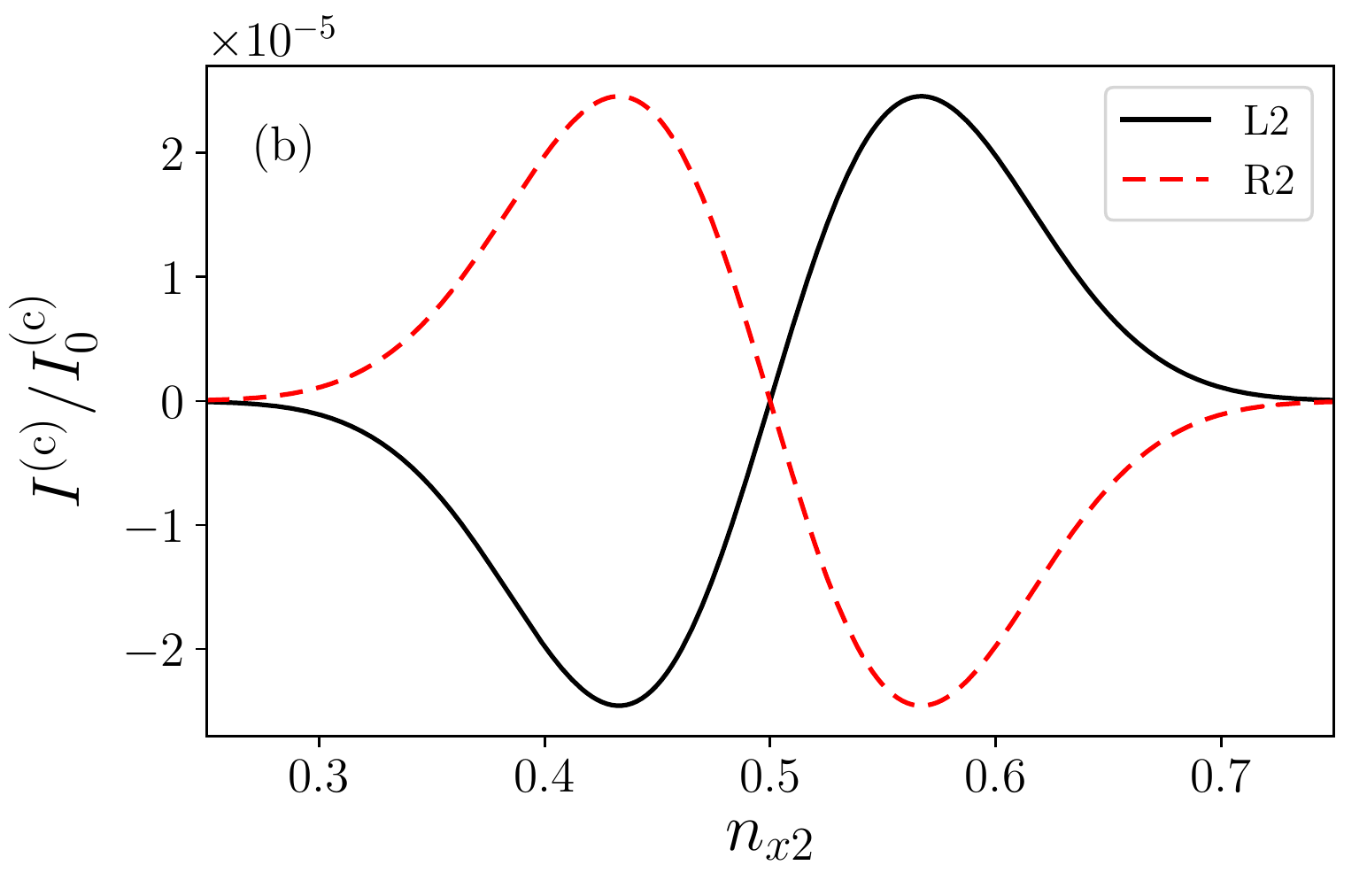}
\caption{
Heat (a) and charge (b) drag currents versus $n_{x2}$ for fixed $n_{x1}=1/2$ in the case where R$2$ is a superconducting electrode.
The other parameters are chosen as follows: ${\cal R}_{\rm L1}= {\cal R}_{\rm R1}={\cal R}_{\rm R2}={\cal R}= 5{\cal R}_{\rm Q}$, ${\cal R}_{\rm L2}=200{\cal R}_{\rm Q}$, $E_{\rm I}=0.3E_C$, $k_{B}T=0.05 E_C$, $\Delta =0.4E_C$.
The charge current is given in units of $I^{\rm (c)}_0=e/(2C{\cal R})$. Charge drag current (b) is obtained by applying a thermal bias $k_{B}\Delta T=0.08 E_C$, whereas the heat drag current (a) is obtained by applying a voltage bias $\Delta V =0.08 E_{\rm C}/e$.
}
\label{fig:dragsc}
\end{figure}

\begin{figure}[ht]
\includegraphics[width=1\columnwidth,clip=true]{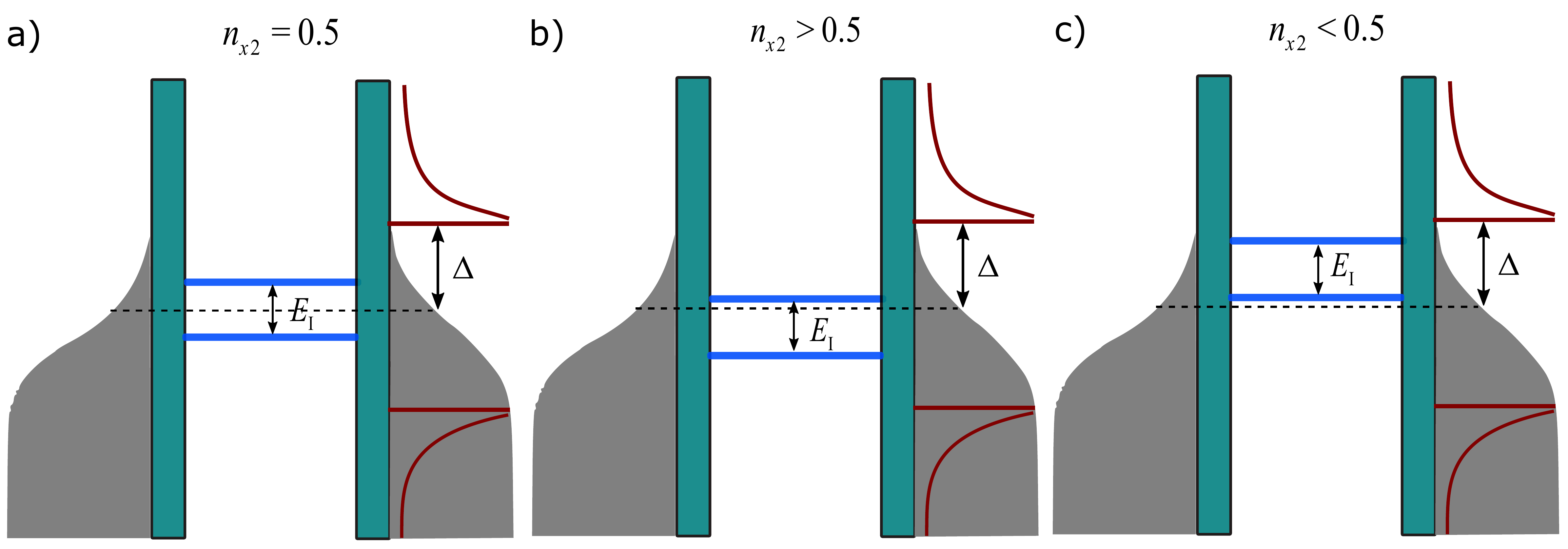}
\caption{Sketch of the energies in the presence of a superconducting electrode on R2, for $n_{x1}=1/2$. The red line represent the superconducting DOS, with a gap equal to $\Delta$ centered at the equilibrium electrochemical potential of the electrodes (dashed thin line). Blue lines represent the two chemical potentials of the lower island, $\delta U_2(1,0)$ and $\delta U_2(0,0)$. Such chemical potentials, according to Eqs.~(\ref{u200}) and (\ref{u210}), are symmetric with respect to the electrochemical potential of the electrodes when $n_{x2}=0.5$ [(panel a)], shift downwards [(panel b)] when $n_{x2}>0.5$, and shift upwards [(panel c)] when $n_{x2}<0.5$.}
\label{energiesS}
\end{figure}

As shown in Fig.~\ref{fig:dragsc}(a), the heat currents in the drag circuit (due to a voltage bias in the drive circuit) plotted as a function of $n_{x2}$, with $n_{x1}=1/2$, exhibit a qualitatively different behavior when compared with the energy-independent case.
Namely, $I^{\rm (h)}_{\rm R2}$ [dashed red curve in Fig.~\ref{fig:dragsc}(a)] is not a bell-shape function, but rather presents two maxima, symmetric with respect to $n_{x2}=1/2$, separated by a shallow dip. This behaviour is a result of the peculiar energy-dependence of the DOS of the superconductor, which presents a gap around the equilibrium electrochemical potential and narrow peaks at $\epsilon=\pm\Delta$ [see Eq.~(\ref{dosSC})]. The former, on the one hand, suppresses the transfer of quasi-particles in and out of the right electrode R2 when $n_{x2}\simeq 1/2$, see Fig.~\ref{energiesS}(a), thus producing a dip in $I^{\rm (h)}_{\rm R2}$. The narrow peaks, on the other hand, promote such transfer when $n_{x2}=1/2-\Delta/(2E_{\rm C})+E_{\rm I}/(4E_{\rm C})$, i.~e. when $n_{x2}$ is such that $\delta U_2(0,0)$ $[\delta U_2(1,0)]$ is close to $\mp\Delta$ [see the sketch in Fig.~\ref{energiesS}(c)], inducing an enhancement of the heat flow into the electrode R2.

Let us now consider the behavior of  $I^{\rm (h)}_{\rm L2}$, represented by the solid black curve in Fig.~\ref{fig:dragsc}(a).
Remarkably,  $I^{\rm (h)}_{\rm L2}$ takes negative values for $n_{x2} \simeq 0.6$ and $n_{x2} \simeq 0.4$, which means that heat is extracted from reservoir L2.
We observe that such heat extraction is related (occurring roughly at the same values of $n_{x2}$) to the peaks in the heat current entering R2.
One could intuitively imagine that the heat extracted from L2 results from a ``compensation'' of the enhanced heat flow entering R2.
We point out that heat extraction occurs only when three conditions are met, namely when $k_{\rm B}T <  E_{\rm I}$, $E_{\rm I} \approx \Delta$ and ${\cal R}_{\rm L2}$ is larger than the other tunnel resistances.

Furthermore, we find that the superconducting electrode R2 allows a finite thermoelectric drag of charge current.
Fig.~\ref{fig:dragsc}(b) shows the dragged charge thermocurrent (i. e. due to a thermal bias in the drive circuit) plotted as a functions of $n_{x2}$, for a fixed $n_{x1}=1/2$.
The dragged charge thermocurrent, on the one hand, vanishes at $n_{x2}=1/2$ because of the symmetric energy configuration [see Fig.~\ref{energiesS}(a)].
For $n_{x2}<1/2$, however, the two chemical potentials shift up [see Fig.~\ref{energiesS}(c))] so that the up most one matches the peak of the DOS of the superconductor, thus favouring a charge current flowing towards the right, i. e. $I^{\rm (c)}_{\rm R2}$ becomes positive.
For $n_{x2}>1/2$, an analogous argument holds for which the transfer of holes towards the right is favoured when the down most chemical potential matches the peak of the DOS of the superconductor [see Fig.~\ref{energiesS}(b)], so that $I^{\rm (c)}_{\rm R2}$ takes negative values.

We remark that the necessity for energy-dependent lead-island couplings in the drag circuit, in order to obtain a drag of charge, was discussed for single-level QD-based Coulomb-coupled systems in Refs.~\onlinecite{RSanchez2010}, \onlinecite{Kaasbjerg2016}, \onlinecite{Keller2016}, \onlinecite{Bischoff2015} and \onlinecite{Volk2015}, in the presence of a voltage bias.
Energy-dependent couplings were introduced through the dependence on the charge state of the QDs of the transition rates between leads and QD, in Refs.~\onlinecite{RSanchez2010} and \onlinecite{Keller2016}, and through the linear energy dependence of the DOS of graphene in Refs.~\onlinecite{Bischoff2015} and \onlinecite{Volk2015}.
These mechanisms, however, are not realistic for metallic islands.

Finally, we wish to mention that, in the limit of small biases, both $I^{\rm (c)}_{\rm drag}$ and $I^{\rm (h)}_{\rm drag}$ are second order in $V$ or $\Delta T$, independently of the values of $n_{x1}$ and $n_{x2}$  (analogously to what found in Ref.~\onlinecite{RSanchez2010} and \onlinecite{Kaasbjerg2016} for the drag of charge in the biased-voltage case of QD-based systems).
We checked that first order contributions in $V$ or $\Delta T$ appear when an additional superconducting electrode is included in the drive circuit, i. e. when energy-dependent lead-island couplings are present in the drive circuit as well as in the drag circuit (analogously to what found in Ref.~\onlinecite{RSanchez2010} for the drag of charge in the biased-voltage case of QD-based systems).
Furthermore, the heat currents in the drag circuit are proportional to $V$ or $\Delta T$ when energy-dependent couplings are present at least in the drive circuit.

\section{Coulomb-coupled quantum wires}
\label{QW}
\begin{figure}[H]
\includegraphics[width=\columnwidth,clip=true]{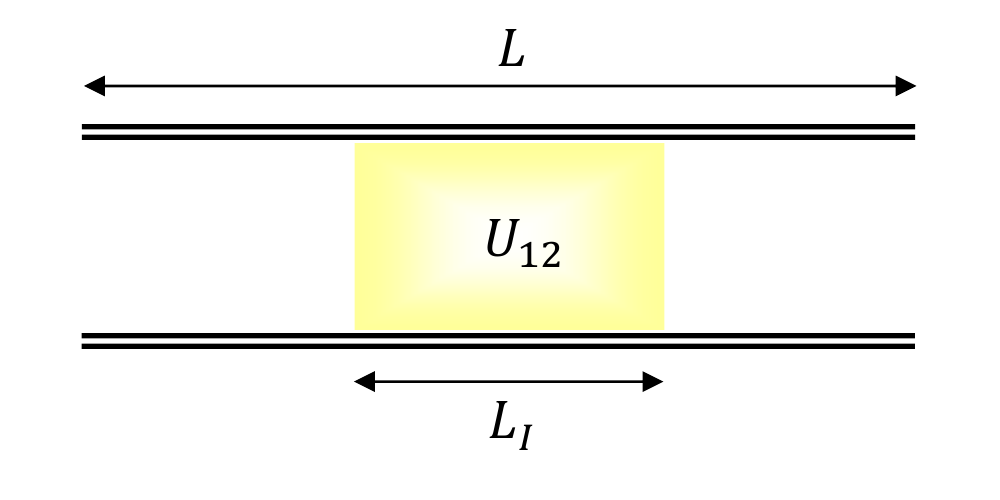}
\caption{Sketch of the two Coulomb-coupled quantum wires of length $L$. The region where the inter-wire interaction is present is $L_{\rm I}$ (with $L_{\rm I}\ll L$) long.}
\label{setupQW}
\end{figure}
In this section we consider the mesoscopic system consisting of two parallel 1D quantum wires (QW) of length $L$ interacting in a region of length $L_{\rm I}\ll L$, see Fig.~\ref{setupQW}, oriented along the $x$-direction.
The total Hamiltonian of the system can be written as
\begin{equation}
H =\int dx {\cal H}_1(x) +\int dx {\cal H}_2(x)+H_{\rm int}
\end{equation}
where ${\cal H}_i(x)$ is the Hamiltonian density of the wire $i=1,2$. At low temperatures ($T\ll T_F$, where $T_F$ is the Fermi temperature), a QW can be described through the Luttinger liquid Hamiltonian,
whose density is
\begin{equation}
{\cal H}_i(x) =\frac{v_i}{2}\left[ gP_i(x)^2+\frac{1}{g}\left( \partial_x \phi_i (x) \right)^2 \right] ,
\end{equation}
where $v_i$ and $g_i$, respectively, are the Luttinger velocity and interaction parameter ($g_i\gtrless1$ for attractive/repulsive intra-wire interaction),
while $\phi_i$ and $P_i$ are canonically conjugated bosonic fields that describe the electronic excitations near the Fermi surface. Details of the bosonization
formalism can be found, for example, in Ref.~\onlinecite{Giam:book}. The coupling between the QWs is
\begin{equation}\label{couplingH}
H_{\rm int}= \int dx dy \;U_{12} (x-y)\rho_1(x) \rho_2(y) ,
\end{equation}
where $\rho_{1/2}$ is the density in  QW 1/2 and $U_{12}$ is the inter-wire Coulomb coupling. $U_{12}$ is assumed to be relevant only in the interaction region of size $L_{\rm I}$.

In the case of the Coulomb drag, the total Hamiltonian commutes with the charge on each wire, so that there cannot be an electric current between the two wires.
This is not the case for the thermal drag, since the coupling does not conserve the energy on each wire and enables an inter-wire energy
transfer. In general, there will be a heat current flowing between the two wires unless they have equal energy.
In addition, if there is a temperature
difference between the extremes of wire 1, there will be a drag heat current flowing in wire 2 as well as a drive current in wire 1. Since there can be heat
transfer between the two wires, the thermal currents at the two extremes of a wire may be different.

We identify with $I^{\rm (c/h)}_\alpha$ the charge/heat current evaluated at the extremes of the QW, i.e. very far from the region of interaction.
We calculate them through the continuity equations expressing the conservation of particle and energy, respectively given by
\begin{equation}
\partial_xI_{\alpha}^{\rm c}+e\partial_t\rho_{\alpha}=0
\end{equation}
and
\begin{equation}
\partial_xI_{\alpha}^{\rm h}+\partial_t\mathcal{H}_{\alpha}=0 .
\end{equation}
In addition to the drag current, Eq.~(\ref{drag}), we define the longitudinal currents in the drive circuit as~\cite{Langer1962}
\begin{equation}
\label{driveheat}
I^{\rm (c/h)}_{\parallel}=\frac{I_{R1}^{\rm (c/h)}-I_{L1}^{\rm (c/h)}}{2} .
\end{equation}
By combining the continuity equations with the Schr\"odinger equation, under the assumption that the region of interaction is much smaller than the size
$L$ of the QWs (and does not scale with $L$), we can write the dragged and longitudinal currents in the bosonized form as
\begin{equation}
\label{dragJbos1}
I^{\rm (h)}_{\rm drag}=-\frac{v_2^2}{2}\frac1L\int dx\{\partial_x\phi_2(x),P_2(x)\}
\end{equation}
and
\begin{equation}
\label{dragJbos2}
I^{\rm (h)}_{\parallel}=-\frac{v_1^2}{2}\frac1L\int dx\{\partial_x\phi_1(x),P_1(x)\} ,
\end{equation}
respectively.

In the linear response regime, we can describe the heat transport using the resistivity matrix $[\rho^{\rm (h)}]$ through the expression
\begin{equation}
\begin{pmatrix}
\nabla T_1 \\
\nabla T_2
\end{pmatrix}=
-[\rho^{\rm (h)}]\begin{pmatrix}
I^{\rm (h)}_{\parallel} \\
I^{\rm (h)}_{\rm drag}
\end{pmatrix},
\end{equation}
where $\nabla T_i$ is the temperature gradient in QW $i$ and
\begin{equation}\label{resistivity}
[\rho^{\rm (h)}]
=-\begin{pmatrix}
\rho_{11}^{\rm (h)} & -\rho_{12}^{\rm (h)} \\
-\rho_{21}^{\rm (h)} & \rho_{22}^{\rm (h)}
\end{pmatrix} .
\end{equation}
In particular, the trans-resistivity is defined as $\rho_{12}^{\rm (h)}\equiv \frac{\nabla T_1}{I_{\rm drag}^{\rm (h)}}$ when $I_{\parallel}^{\rm (h)}=0$.
We calculate $\rho_{12}^{\rm (h)}$ by generalising the Kubo formula to the conductivity matrix (the inverse of $[\rho^{\rm (h)}]$) and then applying the memory function formalism \cite{Zheng:Mac} obtaining
\begin{equation}
\label{rho12hmemory}
\rho_{12}^{\rm (h)}=-\frac{9}{\pi^2}L\int\displaylimits_0^{\infty}dt\int\displaylimits_0^{1/k_BT}
d\beta'\frac{\langle\dot{I}_{\rm drag}^{\rm (h)}(-t)\dot{I}_{\parallel}^{\rm (h)}(i\beta')\rangle}{v_1v_2k_B^5T^4} .
\end{equation}

Note that usually the Kubo formula for the conductivity involves the expectation value of two currents.
When we invert it to find the resistivity, therefore, we ``pay the price'' of having an expectation value of two current derivatives.
However, this actually simplifies the calculations, since for weak
coupling we expand such derivatives in powers of $U_{12}$ and find that the leading order for $\rho_{12}^{\rm (h)}$ is quadratic.
Using
Eqs.~(\ref{couplingH}), (\ref{dragJbos1}), (\ref{dragJbos2}), (\ref{rho12hmemory}) and
\begin{equation}
\dot{I}^{\rm (h)}_{\parallel,\rm drag}=i\int dxdy\; U_{12}(x-y)[\rho_1(x)\rho_2(y),I^{\rm (h)}_{\parallel,\rm drag}] ,
\end{equation}
we switch to the Fourier space and find
\begin{equation}
\label{transH}
\rho_{12}^{\rm (h)}=\int\displaylimits_{k,\omega>0}dkd\omega \frac{9v_1v_2k^2U^2_{12}(k)}{2\pi^4k_B^5T^4}\frac{A_1(k,\omega)
A_2(k,\omega)}{\sinh^2(\omega/2T)} ,
\end{equation}
where
\begin{equation}\label{spectral}
A_i(k,\omega)=\frac12\int dxdt\; e^{-ikx+i\omega t}\langle[\rho_i(x,t),\rho_i(0,0)]\rangle
\end{equation}
is the spectral function of wire $i$.

Let us discuss the physical meaning of Eq.~(\ref{transH}).
The trans-resistivity is quadratic in $U_{12}$ in the limit of weak coupling and it is also proportional to an integral over frequencies of the density-density correlation functions of the two wires, evaluated at equilibrium for $U_{12}=0$.
The integral is weighted by an hyperbolic sine squared, which reflects the bosonic character of the excitations carrying the thermal
current.  As we will discuss below, the dominant contribution to the trans-resistivity is positive: this is not surprising since the moving carriers in wire 2 tend to drag the carriers in wire 1 along their direction of motion.
To keep them at rest a thermal gradient whose direction is the same of $I^{\rm (h)}_{\rm drag}$ must be applied to wire 1.

We notice that Eq.~(\ref{transH}) is identical to the formula for the electrical trans-resistivity~\cite{Zheng:Mac,Pustilnik2003}, except for the factor in
front of the integral.
Indeed, the electrical trans-resistivity $\rho_{12}^{\rm (c)}$ (defined as ${\cal E}_{\rm app}/I^{\rm (c)}_{\rm{drag}}$ at $I^{\rm (c)}_{||}=0$, with ${\cal E}_{\rm app}$ being the electric field applied to the drive wire) is given by
\begin{equation}
\label{transC}
\rho_{12}^{\rm (c)}=\int\displaylimits_{k,\omega>0}dkd\omega \frac{k^2U^2_{12}(k)}{2\pi^2e^2n_1n_2T}
\frac{A_1(k,\omega)A_2(k,\omega)}{\sinh^2(\omega/2T)}
\end{equation}
so that one obtains
\begin{equation}
\label{WF}
\frac{\rho_{12}^{\rm (c)}}{\rho_{12}^{\rm (h)}}=\frac{\pi^2}{9}
\frac{k_B^5T^3}{e^2n_1n_2v_1v_2} .
\end{equation}
This represents a Wiedemann-Franz-like law for drag, which, remarkably, states that the ratio of electrical to thermal trans-resistivity is proportional to $T^3$, in contrast to the ordinary Wiedemann-Franz law for drive currents \cite{KaneFisher} which exhibits a ratio proportional to $T$. This result may be a consequence of the
linear spectrum approximation made in the limit of low temperatures: in this limit the thermal current is linear in $k$, exactly as the electric current.

For the sake of definiteness, we will now assume that the inter-wire coupling has the following specific form,
\begin{equation}
U_{12}(k) = U_{12}(0) e^{-k/k_0} .
\end{equation}
Although the quantitative behaviour of the thermal trans-resistivity depends on this choice, we expect the qualitative features to be far more general.
The relevant temperature scales are: i) $T_0=v_{F}k_0/k_B$, associated to the typical wavevector scale $k_0$ over which the coupling
$U_{12}(k)$ decays  ($k_0$ is very small for long range interactions, while approaches infinity for point-like interactions); ii) $T_1= k_{F}\delta v/k_B$,
associated to the difference between the Luttinger velocities of the two wires $\delta v=v_2-v_1$, assumed to be small, so that $T_1\ll T_0$.
In the definition of $T_0$, we have used the average Fermi velocity $v_F=v g$, where $v=(v_1+v_2)/2$ and $g=(g_1+g_2)/2$.

\begin{figure}[ht]
\includegraphics[width=1\columnwidth,clip=true]{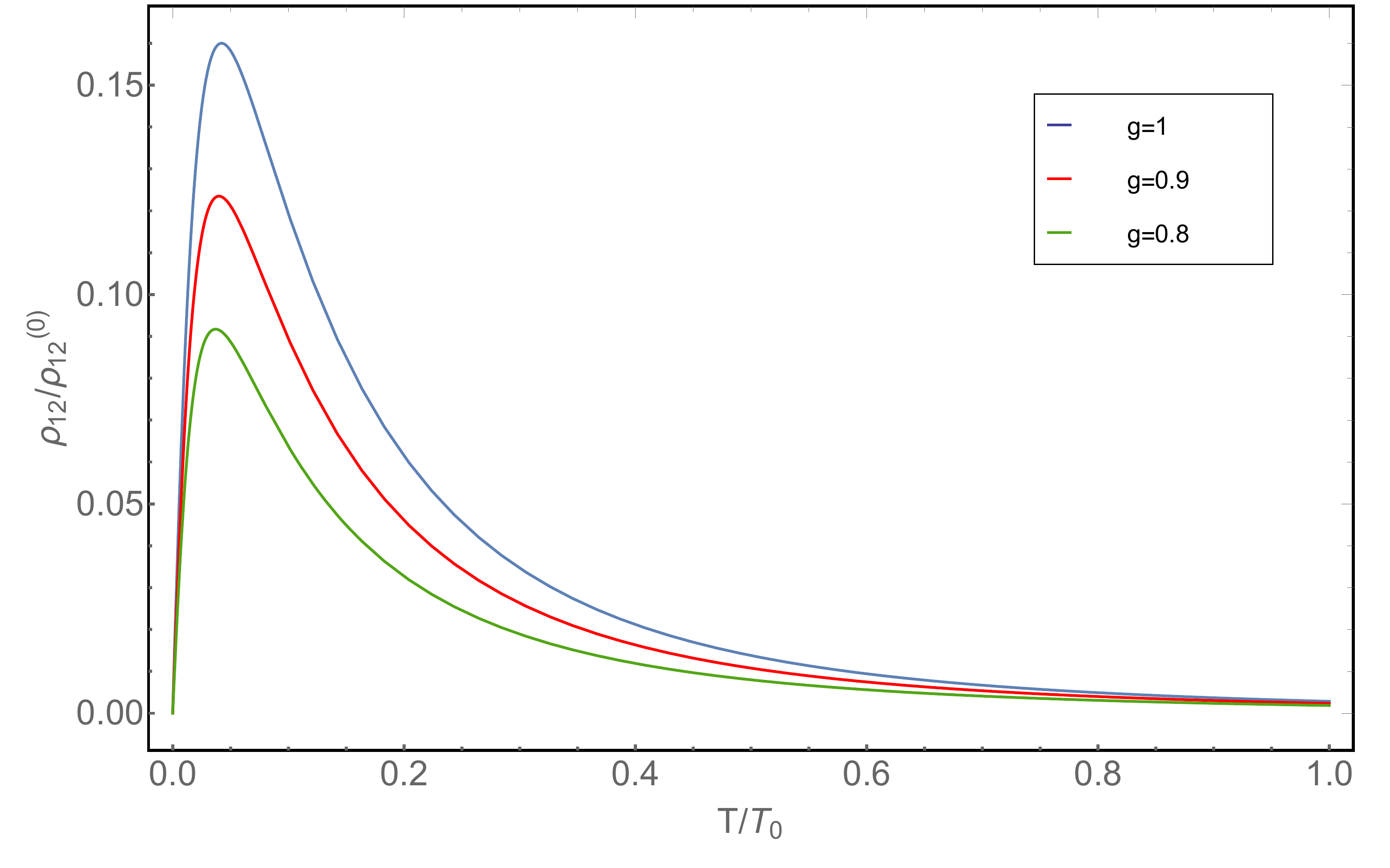}
\vspace*{0.25cm}
\includegraphics[width=1\columnwidth,clip=true]{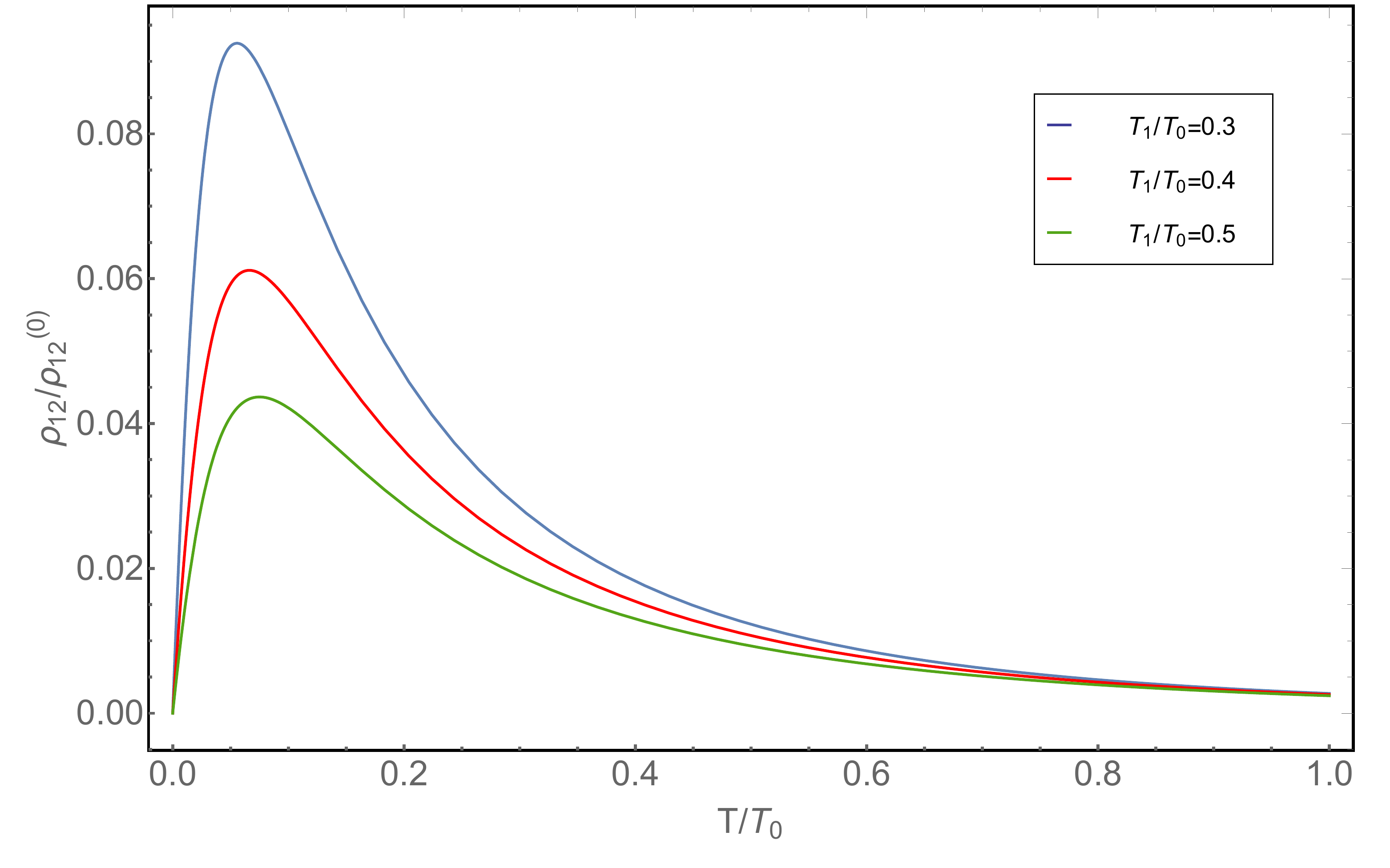}
\caption{(Color online). Plot of $\rho_{12}^{\rm (h)}/\rho_{12}^{(0)}$ for various values of $g$ and fixed $T_1/T_0=0.2$ (top panel), and for various values of $T_1$ and fixed $g=1$ (bottom panel). We have defined $\rho_{12}^{(0)}\equiv mU^2(0)/(v_Fk_B^2T_0)$.}
\label{rhogT&TT}
\end{figure}

\begin{figure}[ht]
\includegraphics[width=1\columnwidth,clip=true]{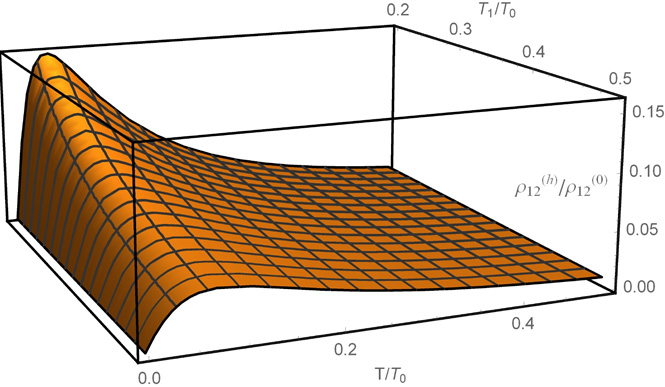}
\caption{3D plot of $\rho_{12}^{\rm (h)}/\rho_{12}^{(0)}$ as a function of $T/T_0$ and $T_1/T_0$ for $g=1$.}
\label{rho3D}
\end{figure}

$A_i(k,\omega)$ has two contributions: one coming from the forward scattering (in which a small momentum transfer near the Fermi surface occurs)
and one originating from the back scattering, involving a momentum transfer between the opposite sides of the Fermi surface.
The second contribution is negligible for $T\ll T_F$ and it is not taken into account any further.
For well defined excitation we would have $A_i(k,\omega)\sim\delta(\omega-v_ik)$.
However, we need to consider the
finite lifetime of these excitations, such that $A_i$ has a certain width and it is given by
\begin{equation}\label{Ai}
A_i(k,\omega)\sim\frac{\delta\omega}{(\omega-v_ik)^2+\delta\omega^2},
\end{equation}
where $\delta\omega$ is a quantity related to the band dispersion curvature and to the thermal broadening of excitations \cite{Pustilnik2003}. $A_1$ and $A_2$ are peaked at different
positions ($v_1k$ and $v_2k$, respectively), so that their overlap is determined by $T_1$.
In fact, the smaller $T_1$ is, the larger the overlap and $\rho_{12}^{\rm (h)}$ are.
We follow a procedure similar to the one outlined in Ref.~[\onlinecite{Pustilnik2003}] to calculate the integral in
Eq.~(\ref{transH}) and find that $\rho_{12}^{\rm (h)}\sim T$ for $T\ll T_1$, $\rho_{12}^{\rm (h)}\sim1/T$ for $T_1\ll T\ll T_0$ and
$\rho_{12}^{\rm (h)}\sim1/T^3$ for $T_0\ll T$.
As a result, the trans-resistivity presents a peak for intermediate temperatures and goes to zero for low
and high $T$.
This is shown in Fig.~\ref{rhogT&TT}, where the normalized thermal trans-resistivity is plotted as a function of temperature $T$ for various values of $g$ and fixed $T_1=0.2T_0$ (top panel) and for various values of $T_1$ and fixed $g=1$ (bottom panel).
In particular, Fig.~\ref{rhogT&TT} (top panel) shows that $\rho_{12}^{\rm (h)}$ gets suppressed for stronger repulsive interaction (or smaller $g$).
For completeness, we report in Fig.~\ref{rho3D} the 3D plot of the normalized $\rho_{12}^{\rm (h)}$ as a function of $T$ and the ratio $T_1/T_0$.
We stress that $\rho_{12}^{\rm (h)}$ turns out to be proportional to $U(0)^2$, the strength of inter-wire coupling at $k=0$.
Finally, by defining the trans-conductivity as $\sigma_{21}^{\rm (h)}\equiv -J_{\rm drag}^{\rm (h)}/\nabla T_1$ (at $\nabla T_2=0$) we find that
$\sigma_{21}^{\rm (h)}\propto T^2\rho_{12}^{\rm (h)}$, so that its peak is shifted to larger $T$ compared to the peak of $\rho_{12}^{\rm (h)}$.

\subsection{Numerical simulations}
\label{numsim}
\begin{figure}[t]
\includegraphics[width=1\columnwidth,clip=true]{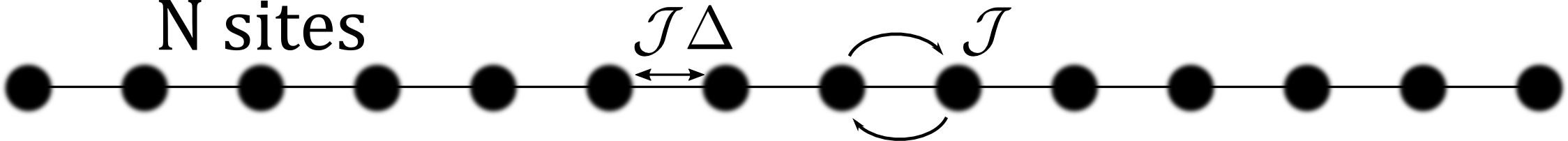}
\caption{Diagram of the discrete system used with next neighbor hopping (${\cal J}$) and interaction (${\cal J}\Delta$) terms.}
\label{spinchain}
\end{figure}

In this section, we complement our analytical study with numerical computations. Namely we model each quantum wire with a discrete chain of spinless electrons (Fig.~\ref{spinchain}) and employ a protocol based on the matrix product states (MPS) formalism \cite{Sch:wol} to simulate the time evolution of our system out of equilibrium.
When a temperature gradient is present, the current reaches a non equilibrium steady state after a transient phase~\cite{Maz:ros,Karrasch}. We study the system at temperatures comparable with the Fermi energy and focus on the dependence on the inter-wire interaction, exploring the regime of strong couplings.

The Hamiltonian of the electrons chain is related to the spin 1/2 XXZ model by a Wigner-Jordan transformation \cite{Giam:book}.
The kinetic ($H_{\rm kin}$) and interaction ($H_{\rm int}$) terms are modelled in a simple way, using a next neighbour hopping and a next neighbour interaction as follows
\begin{gather}\label{Hchain}
H_{\rm kin}=-\frac{{\cal J}}{2} \sum_i \left( a_{i+1}^\dagger a_{i}+a_{i}^\dagger a_{i+1} \right) ,\\
H_{\rm int}={\cal J} \Delta \sum_i \left( a_{i+1}^\dagger a_{i+1}- \frac{1}{2} \right)\label{Hchain2}
\left( a_{i}^\dagger a_{i}- \frac{1}{2} \right) .
\end{gather}
Here $i=1,...,N$ labels the electron sites, while $a_{i}^\dagger$ ($a_{i}$) create (annihilate) an electron on site $i$. The hopping parameter $\cal{J}$ has the dimensions of an energy and is related to the Fermi velocity of the system.
The next neighbor interaction depends on the densities of the two neighboring sites and its strength is tuned by the dimensionless parameter $\Delta$, which determines the phase of the system. In fact, for $\Delta<-1$ the chain is in a ferromagnetic ordered phase, while $\Delta>1$ corresponds to an antiferromagnetic one.
For $|\Delta|<1$ the Hamiltonian describes a system of interacting electrons with either repulsive ($\Delta>0$) or attractive ($\Delta<0$) interaction. At low temperatures Eqs.~(\ref{Hchain}-\ref{Hchain2}) describe a Luttinger liquid and there are precise relations between $\cal J$ and $\Delta$ and the Luttinger parameters $v$ and $g$~\cite{Giam:book}.

Once the Hamiltonian for one wire is defined, we just need to consider two of them and model the inter-wire interaction as a local coupling.
More precisely, we assume that the inter-wire interaction is proportional to the densities and non-zero only for corresponding sites on the two QWs, i. e.
\begin{equation}
H_{\rm 12}=\sum_i U_i \left( a_{i,1}^\dagger a_{i,1}- \frac{1}{2} \right)
\left( a_{i,2}^\dagger a_{i,2}- \frac{1}{2} \right) ,
\end{equation}
where $U_i$ represents the interwire coupling.
For the sake of simplicity, we choose $U_i=U$ for the two couples of central sites and $U_i=0$ otherwise.
Therefore, the total Hamiltonian $H$ is simply the sum $H=H_{\rm kin,1}+H_{\rm int,1}+H_{\rm kin,2}+H_{\rm int,2}+H_{12}$.
\begin{figure}[t]
\includegraphics[width=1\columnwidth,clip=true]{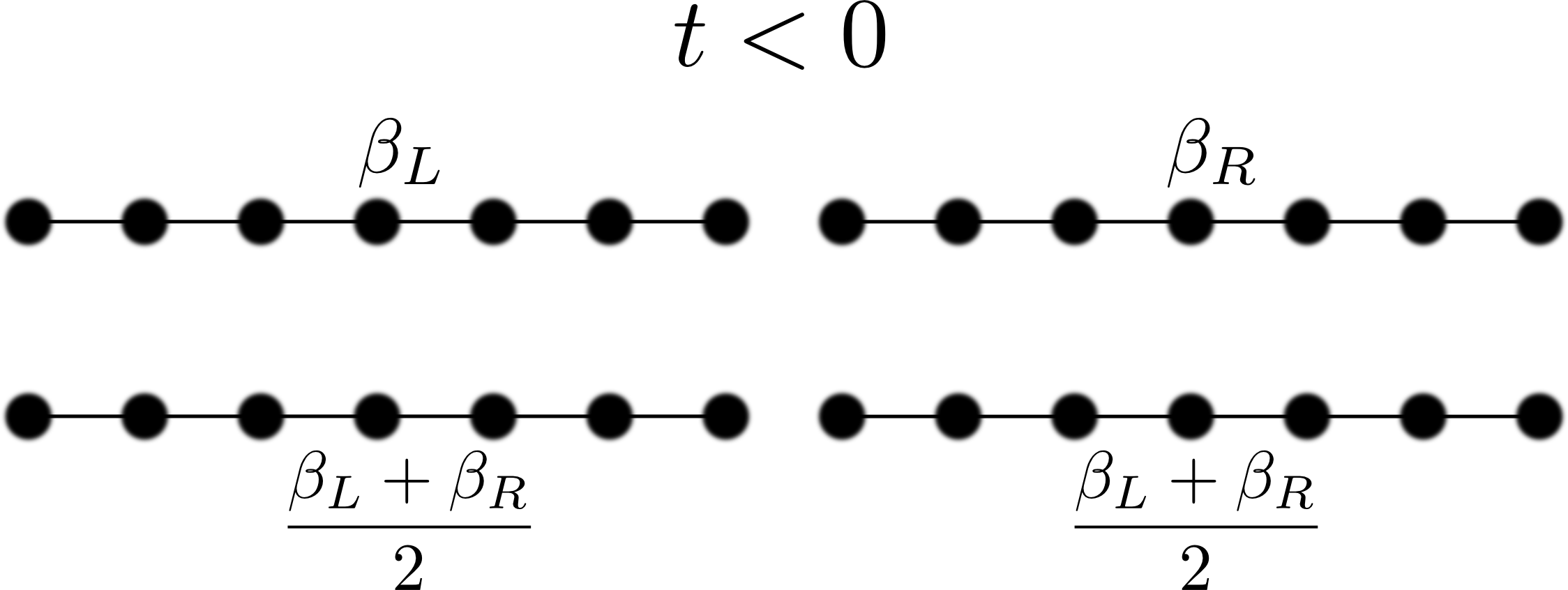} \\
\noindent\rule{\columnwidth}{0.4pt} \\
\vspace*{0.15cm}
\includegraphics[width=1\columnwidth,clip=true]{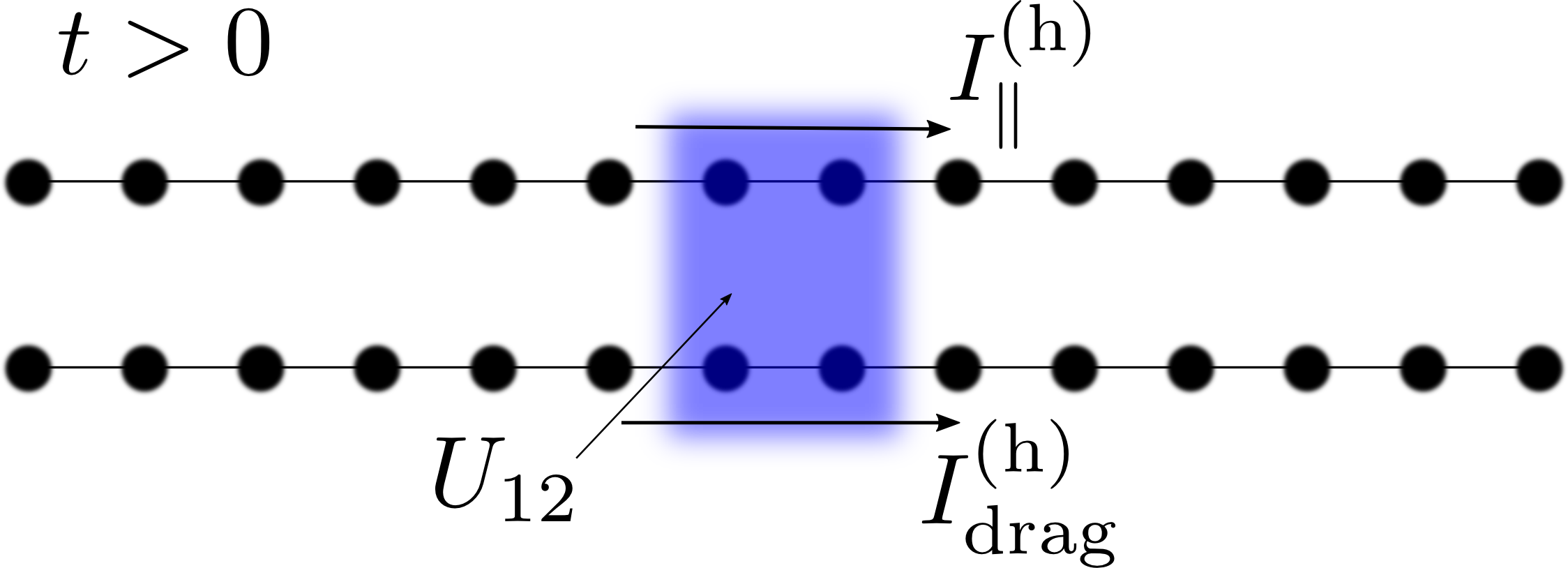}
\caption{Schematic picture of the thermal state of the system for $t\leq0$ (top panel) and of the time evolution of the system for $t>0$ (bottom panel). In the upper panel, for each half of the wire the inverse temperature is shown. In the lower panel, the coupling region is coloured in blue and the drive and drag currents are shown.}
\label{timev}
\end{figure}

The MPS formalism enables to perform a time evolution (both real and imaginary) of the system, thus allowing to assign a certain temperature to the system and a temperature gradient to the drive QW by evolving them into the appropriate thermal state. The real time evolution then allows to reach the steady state.
The detailed protocol used in the computations follows two main steps (see Fig.~\ref{timev}):
\begin{itemize}
  \item The system is assigned into its thermal state using an imaginary time evolution. This evolution is carried out without considering the inter-wire coupling and removing the next neighbor interaction between the central couples of sites, so that the wires are effectively split into halves. The left half of the drive wire is evolved into a state with inverse temperature $\beta_L$, while the right half is evolved up to a different $\beta_R$ (it is the crudest yet simplest way to create a temperature difference). In the drag chain, we chose the left and right half to have the same inverse temperature $(\beta_L+\beta_R)/2$. For each half of the wires we use the corresponding Hamiltonian to perform the imaginary time evolution.
  \item At $t=0$ the next neighbors interaction between the central sites is switched on, connecting the halves of the two wires, and we also turn on the inter-wire coupling. For $t>0$ the system is now evolved in real time using the complete hamiltonian $H$. After a transient phase, the system and the drag current reach a stationary state.
\end{itemize}

In the numerical simulations we can tune the following parameters: $U$, $\Delta$, $N$, $\beta_L$ and $\beta_R$.
Because of computational time constraints, we choose the number of sites to be $N=20$ (we check that this value is high enough to ensure that the finite size effects are not important).
Notice that on the relevant time scales, the energy of the sites at the edges of the wires does not change appreciably, meaning that the system has not thermalized yet.
Since we can not choose small values of temperature, as the computational time would be too long, we select values of $\beta_L$ and $\beta_R$ of order $\mathcal{J}^{-1}$.
The simulations are run with various values of $\Delta$ in the range $-1<\Delta<1$, though we observe that the differences are mostly quantitative and thus focus mainly on the free electrons case ($\Delta=0$).
The heat currents $I^{\rm (h)}_{\rm L2}$ and $I^{\rm (h)}_{\rm R2}$ are calculated as discrete time derivatives of the energy in the left/right half of the drag wire.
By observing the time-dependent behavior of the drag current $I^{\rm (h)}_{\rm{drag}}$, its stationary value can be extracted within a proper time-window (see App.~\ref{QWtd} for details).

\begin{figure}[ht]
\hspace*{-0.05\columnwidth}
\includegraphics[width=0.96\columnwidth,clip=true]{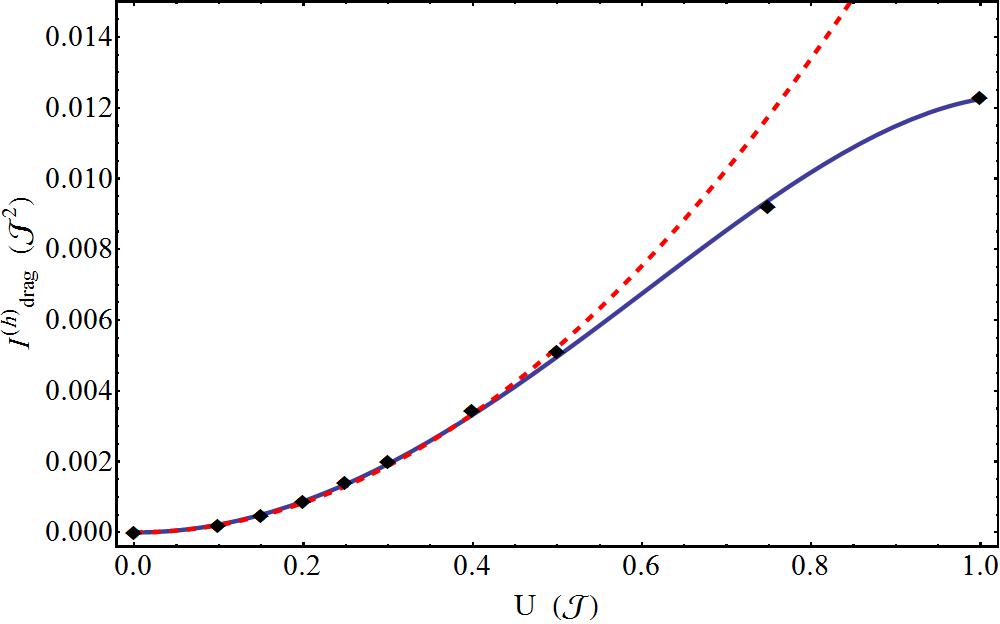}
\caption{Plot of the stationary value of $I^{\rm (h)}_{\rm{drag}}$ as a function of $U$ for $\Delta=0$, $\beta_L=0.5\mathcal{J}^{-1}$ and $\beta_R=0.75\mathcal{J}^{-1}$. The square points represent the data from the simulations, the blue solid line ($aU^2+bU^4$) fits all the points, while the red dotted line (parabola) misses the last two points (for $U\gtrsim0.5\cal J$).}
\label{J2U2}
\end{figure}

We now focus on the dependence on $U$ of the stationary $I^{\rm (h)}_{\rm{drag}}$, which we plot in Fig.~\ref{J2U2} for $\Delta=0$, $\beta_L=0.5\mathcal{J}^{-1}$, $\beta_R=0.75\mathcal{J}^{-1}$ and positive values of $U$ (inversion symmetry holds when changing the sign of $U$).
We notice that the behavior of the drag current is quadratic in $U$ for weak couplings up to $U\sim0.5\cal J$.
The data points from the numerical simulation (represented as black squares in Fig.~\ref{J2U2}) are fitted with a parabola (dotted red line) and with a polynomial curve of the form $aU^2+bU^4$ (blue solid line).
We find that the quartic correction is negative (i. e. $b<0$) and it is in very good agreement with the numerical data up to $U\sim\cal J$.
Notice that the results of the simulations extend the $U^2$-dependence of $I^{\rm (h)}_{\rm{drag}}$, found in the previous section for low temperatures and weak coupling, to a larger temperature range.
\section{Conclusions}
\label{conclusions}
In summary, in this paper we have studied the phenomenon of electronic thermal drag in two different setups, namely for capacitively-coupled metallic islands and for parallel quantum wires.
In the metallic island case, using the master equation approach we have studied both the sequential and the co-tunneling contributions to thermal drag in the presence of either a voltage bias or a temperature bias.
In the sequential tunneling regime we have obtained analytical results for small biases, finding, in particular, that $I^{\rm (h)}_{\rm drag}$ is quadratic in $\Delta T$ or $V$ and non-monotonous as a function the coupling between the islands (inter-island repulsion).
We have found that such behavior holds even when co-tunneling processes are included.
Finally, we have explored the consequences of energy-dependent island-electrode coupling by replacing one of the electrodes in the drag circuit with a superconductor.
Apart from allowing a finite dragged charge current, we have found that the presence of the superconducting electrode can cause the extraction of heat from the remaining normal electrode in the drag circuit.

In the case of the two interacting parallel quantum wires, we have derived an analytic expression for the thermal trans-resistivity $\rho^{\rm (h)}_{12}$, using the Luttinger liquid theory and the bosonization technique, in the weak-coupling limit and at low temperatures.
We have found that $\rho^{\rm (h)}_{12}$ turns out to be proportional to the electric trans-resistivity $\rho^{\rm (c)}_{12}$, in such a way that their ratio is proportional to $T^3$ and obeys a sort of Wiedemann-Franz law for drag.
Furthermore, we have analyzed the behavior of the thermal trans-resistivity in the temperature ranges defined by two temperature scales which naturally emerge: $T_0$, associated to the characteristic wave-vector of the coupling, and $T_1$ (with $T_1 < T_0$), associated to the difference between the Luttinger velocities of the two wires.
We have found that $\rho^{\rm (h)}_{12}$ behaves linearly in $T$ for $T \ll T_1$, decreases like $1/T$ for $T_1\ll T \ll T_0$ or like $1/T^3$ for $T_0 \ll T$, and presents a peak in between these regimes.
Finally, we have performed numerical simulations that allowed to confirm our analytical results in the weak-coupling regime and to access the strong-coupling regime.
We have showed that, in the latter case, $\rho^{\rm (h)}_{12}$ acquires a quartic correction in the inter-wire coupling which adds up to the quadratic behavior characteristic of the weak-coupling regime.

As argued in the following, both setups are experimentally feasible with current technology. In the case of metallic islands, the Coulomb coupling between two such islands has been realised by placing close together two single electron transistors, while making sure that no electron transfer occurs between them, see for example Refs.~\onlinecite{Koski2015} and \onlinecite{Shilpi2017}. On the other hand, Coulomb coupling between a pair of quantum wires was realised already in Refs.~\onlinecite{Debray2001} and \onlinecite{Yamamoto2002} for the measurement of the Coulomb drag (more experimental literature can be found in the review Ref.~\onlinecite{Narozhny2016}). In both cases, heat currents can be determined by making use of heat budget models which account for all possible heat exchanges between the systems and their environment. See, for example, Ref.~\onlinecite{Dutta2017} for the case of metallic islands, and Ref.~\onlinecite{Kim2001}, for the case of quantum wires (multiwalled nanotubes).

We believe that the results obtained in this paper can be also relevant for the implementation of non-local thermal machines. Indeed, the four-terminal system depicted in Fig.~\ref{setup} can be operated as a non-local heat engine where the temperature difference between the two upper reservoirs can be used to extract work from the lower circuit. Likewise, a non-local refrigerator uses the work performed on the upper circuit to cool one of the lower reservoirs. Autonomous refrigerators, where heat is provided instead of work in the upper circuit, can also be envisaged (see Refs.~\onlinecite{Benenti2013} and \onlinecite{Paolo2018}). Moreover, the four-terminal setup can be operated as a thermal gating system, similarly to the three-terminal setups of Refs.~\onlinecite{Thierschmann2015} and \onlinecite{Sanchez2017b}, where the heat or charge flow in the upper circuit is controlled by changing the temperature of the reservoirs in the lower circuit.

\section{Acknowledgments}
We would like to acknowledge fruitful discussions with Jukka Pekola. This work has been supported by SNS-WIS joint lab "QUANTRA", by the
SNS internal projects ÒThermoelectricity in nano-devicesÓ, and ÒNon-equilibrium dynamics of one-dimensional quantum
systems: From synchronisation to many-body localisationÓ, by the CNR-CONICET cooperation programme ÒEnergy conversion in quantum, nanoscale, hybrid devicesÓ and by the COST ActionMP1209 ÒThermodynamics in the quantum regime.Ó

\appendix

\begin{widetext}

\section{Master equations for two capacitively-coupled islands}
\label{ME}
The expressions for the master equations which involve all possible sequential and co-tunneling particle transition rates are
\begin{align}\label{masterequations}
&-\sum_{\alpha\nu}\left[\Gamma_{\nu, 1}^{(c)}(n_{1},n_{2})+\Gamma_{\alpha, 2}^{(c)}(n_{1},n_{2})+\gamma_{\alpha \nu}^{(c)}(n_{1},n_{2})\right] \times p(n_1,n_2) + \sum_{\nu}\Gamma_{1,\nu}^{(c)}(n_{1},n_{2}) \times p(n_1+1,n_2) \nonumber \\
& + \sum_{\alpha}\Gamma_{2,\alpha}^{(c)}(n_{1},n_{2}) \times p(n_1,n_2+1)
+ \sum_{\alpha\nu}\gamma_{\alpha\nu}^{(c)}(n_{1}+1,n_{2}+1) \times p(n_1+1,n_2+1)=0 ,\nonumber \\
&  \sum_{\nu}\Gamma_{\nu, 1}^{(c)}(n_{1},n_{2})\times p(n_1,n_2) - \sum_{\alpha\nu}\left[\Gamma_{1,\nu }^{(c)}(n_{1},n_{2})+\Gamma_{\alpha, 2}^{(c)}(n_{1}+1,n_{2})+\gamma_{\alpha \nu}^{(c)}(n_{1}+1,n_{2})\right]\times p(n_1+1,n_2) \nonumber \\
&+ \sum_{\alpha\nu}\gamma_{\alpha\nu}^{(c)}(n_1,n_{2}+1)\times p(n_1,n_2+1)
+\sum_{\alpha}\Gamma_{2,\alpha}^{(c)}(n_{1}+1,n_{2})\times p(n_1+1,n_2+1)=0, \nonumber \\
&\sum_{\alpha}\Gamma_{\alpha, 2}^{(c)}(n_{1},n_{2})\times p(n_1,n_2) + \sum_{\alpha\nu}\gamma_{\alpha\nu}^{(c)}(n_1+1,n_{2})\times p(n_1+1,n_2) + \sum_{\nu}\Gamma_{1,\nu}^{(c)}(n_{1},n_{2}+1)\times p(n_1+1,n_2+1) \nonumber \\
&-  \sum_{\alpha\nu}\left[\Gamma_{2,\alpha}^{(c)}(n_{1},n_{2})+\Gamma_{\nu, 1}^{(c)}(n_{1},n_{2}+1)+\gamma_{\alpha \nu}^{(c)}(n_{1},n_{2}+1)\right]\times p(n_1,n_2+1)=0, \nonumber \\
&\sum_{\alpha\nu} \gamma_{\alpha \nu}^{(c)}(n_{1},n_{2})\times p(n_1,n_2) + \sum_{\alpha}\Gamma_{\alpha, 2}^{(c)}(n_{1}+1,n_{2})\times p(n_1+1,n_2) + \sum_{\nu}\Gamma_{\nu, 1}^{(c)}(n_{1},n_{2}+1)\times p(n_1,n_2+1) \nonumber \\
&- \sum_{\alpha\nu}\Big[\Gamma_{2,\alpha}^{(c)}(n_{1}+1,n_{2})+\Gamma_{1,\nu}^{(c)}(n_{1},n_{2}+1)
+\gamma_{\alpha \nu}^{(c)}(n_{1}+1,n_{2}+1)\Big]\times p(n_1+1,n_2+1)=0,
\end{align}
where $p(n_{i},n_{j})$ gives the occupation probability for the states $(n_{i},n_{j})$, $\alpha = \{{\rm L}{2},{\rm R}{2}\}$ and $\nu=\{{\rm L}{1},{\rm R}{1}\}$.
The sequential tunneling rates $\Gamma_{\alpha, \nu}^{(c)}(n_{1},n_{2})$ are given by Eqs.~(\ref{gamma1}) and (\ref{gamma2}) whereas the co-tunneling rates $ \gamma_{\alpha \nu}^{(c)}(n_{1},n_{2})$ are given in Appendix~\ref{Rate}.
From the conservation of probability, i.e.
\[p(n_{1},n_2)+p(n_{1}+1,n_2)+p(n_{1},n_2+1)+p(n_{1}+1,n_2+1)=1,\]
the master equations can be solved to obtain the probabilities in terms of transition rates.

\section{Charge and heat transition rates for co-tunneling}
\label{Rate}
Assuming small biases and low temperature, only four states for the occupation of the islands need to be taken into account.
When the initial state is $(n_{1},n_{2})$, such transition rates are associated to an electron reaching island 2 from lead $\alpha=$L$2$, R$2$ and another electron reaching island 1 from lead $\nu={\rm L}1$, R$1$ and can be written as
\begin{equation}
\gamma_{\alpha \nu}^{({\rm c}/{\rm h})}(n_{1},n_{2})=\frac{1}{e^2{\cal R}_{\alpha}}\frac{1}{e^2{\cal R}_{\nu}}H_{n_1,n_2}^{({\rm c}/{\rm h})}\left(\delta U_{2},\delta U_{1}-eV_{\nu},\delta U_{1}+\delta U_{2}+E_{I}-eV_{\nu}\right) ,
\end{equation}
where,
\begin{equation}
H_{n_1,n_2}^{({\rm c}/{\rm h})}\left(E_{1},E_{2}, E_{3}\right)=\frac{\hbar}{2\pi}\int_{-\infty}^{\infty} d\xi {F}_{\alpha 2}^{({\rm c}/{\rm h})}(-\xi) F_{\nu 1}^{(c)}\left(\xi+ E_{3}\right)\left|\frac{1}{\xi+E_{1}-i\eta}-\frac{1}{\xi+E_{3}-E_{2}-i\eta}\right|^{2}
\label{cotfun1}
\end{equation}
is a function whose first and second arguments represent the intermediate energy states due to tunneling in island 2 and island 1, respectively, while the third argument represents the total change in energy of the cotunnelling process.
Similarly, when the initial state is $(n_{1}+1,n_{2}+1)$, the transition rates are associated to an electron reaching island 2 from lead $\alpha=$L$2$, R$2$ and another electron reaching island 1 from lead $\nu=$L$1$, R$1$ and can be written as
\begin{equation}
\gamma_{\alpha \nu}^{({\rm c}/{\rm h})}(n_{1}+1,n_{2}+1)=\frac{1}{e^2{\cal R}_{\alpha}}\frac{1}{e^2{\cal R}_{\nu}}H_{n_1 +1,n_2 +1}^{({\rm c}/{\rm h})}\left(-\delta U_{2}-E_{I},-\delta U_{1}-E_{I}+eV_{\nu},-\delta U_{1}-\delta U_{2}-E_{I}+eV_{\nu}\right) .
\label{cotfun2}
\end{equation}
\begin{equation}
H_{n_1 +1,n_2 +1}^{({\rm c}/{\rm h})}\left(E_{1},E_{2}, E_{3}\right)=\frac{\hbar}{2\pi}\int_{-\infty}^{\infty} d\xi \,{G}_{ 2\alpha}^{({\rm c}/{\rm h})}(\xi) G_{ 1\nu}^{(c)}\left(-\xi- E_{3}\right)\left|\frac{1}{\xi+E_{1}-i\eta}-\frac{1}{\xi+E_{3}-E_{2}-i\eta}\right|^{2}
\end{equation}
Analogously, the expressions relative the remaining initial states $(n_{1}+1,n_{2})$ and $(n_{1},n_{2}+1)$ can be written as
\begin{equation}
\gamma_{\alpha \nu}^{({\rm c}/{\rm h})}(n_{1}+1,n_{2})=\frac{1}{e^2{\cal R}_{\alpha}}\frac{1}{e^2{\cal R}_{\nu}}H_{n_1 +1,n_2}^{({\rm c}/{\rm h})}\left(\delta U_{2}+E_{I},-\delta U_{1}+eV_{\nu},-\delta U_{1}+\delta U_{2}+eV_{\nu}\right)
\label{cotfun3}
\end{equation}
and
\begin{equation}
\gamma_{\alpha \nu}^{({\rm c}/{\rm h})}(n_{1},n_{2}+1)=\frac{1}{e^2{\cal R}_{\alpha}}\frac{1}{e^2{\cal R}_{\nu}}H_{n_1,n_2 +1}^{({\rm c}/{\rm h})}\left(-\delta U_{2},\delta U_{1}+E_{I}-eV_{\nu},\delta U_{1}-\delta U_{2}-eV_{\nu}\right) ,
\label{cotfun4}
\end{equation}
respectively, where
\begin{equation}
H_{n_1 + 1, n_2 }^{({\rm c}/{\rm h})}\left(E_{1},E_{2}, E_{3}\right)=\frac{\hbar}{2\pi}\int_{-\infty}^{\infty} d\xi {F}_{\alpha 2}^{({\rm c}/{\rm h})}(-\xi) G_{1 \nu}^{(c)}\left(-\xi - E_{3}\right)\left|\frac{1}{\xi+E_{1}-i\eta}-\frac{1}{\xi+E_{3}-E_{2}-i\eta}\right|^{2} .
\label{cotfun}
\end{equation}
\begin{equation}
H_{n_1 , n_2 +1}^{({\rm c}/{\rm h})}\left(E_{1},E_{2}, E_{3}\right)=\frac{\hbar}{2\pi}\int_{-\infty}^{\infty} d\xi {G}_{ 2\alpha}^{({\rm c}/{\rm h})}(\xi) F_{ \nu 1}^{(c)}\left(\xi + E_{3}\right)\left|\frac{1}{\xi+E_{1}-i\eta}-\frac{1}{\xi+E_{3}-E_{2}-i\eta}\right|^{2} .
\end{equation}
We could compute the integrals in Eqs.~(\ref{cotfun1}) and (\ref{cotfun}) exactly at first order in $\Delta T$ and for arbitrary $V$ using the standard approaches\cite{Turek2002,Ruokola2011,Kaasbjerg2016} (see App.~\ref{Curr}).
Note, in particular, that the master equations are affected by the co-tunneling processes that involve electrodes on different circuites, since they change the occupations of the islands (see App.~\ref{ME}).

\subsection{Derivation}
The co-tunneling rate in its most general form when one electron enters island 1 and another electron leaves island 2 can be written as:
\begin{align}
\gamma_{\alpha \nu}^{({\rm c})}=&\frac{\hbar}{2\pi e^{4}R_{\alpha}R_{\nu}}\int d E_{\alpha}d E_{\nu}d E_{s_{1}}d E_{s_{2}}f_{\nu}(E_{\nu})\left(1-f_{\alpha}(E_{\alpha})\right)f_{2}(E_{s_{2}})\left(1-f_{1}(E_{s_{1}})\right)\nonumber \\
&\left|\frac{1}{E^{+}+E_{s_{1}}-E_{\nu}}+\frac{1}{E^{-}+E_{\alpha}-E_{s_{2}}}\right|^{2}\delta \left(E_{\alpha}-E_{\nu}+E_{s_{1}}-E_{s_{2}}+\Delta E\right)\nonumber \\
&=\frac{\hbar}{2\pi e^{4}R_{\alpha}R_{\nu}}\int  d \xi_{\alpha} d\xi_{\nu}F_{\nu 1}^{(c)}(-\xi_{\nu})G_{2\alpha}^{(c)}(\xi_{\alpha})\left|\frac{1}{E^{+}+\xi_{\nu}}+\frac{1}{E^{-}+\xi_{\alpha}}\right|^{2}\delta\left(\xi_{\alpha}+\xi_{\nu}+\Delta E\right)
\label{cot}
\end{align}
where, $E_{j}$, $j=\left\{\alpha, \nu , s_i\right\}$ refers to the energy states of the corresponding leads and islands; $\Delta E$ is the total energy change in the co-tunneling process which can be written in terms of change in electrostatic energy and potential bias. Since the temperature of both of the islands is same, the above expression for co-tunneling rate is equally applicable for co-tunneling involving only one island. Also, $G_{2\alpha}^{({\rm c})}$ is the sequential tunneling rate corresponding to the $\alpha^{\text{th}}$ reservoir given by Eq. (\ref{fu2}).
\subsection*{Case 1: $\mathbf{\Delta T =0}$}
When $\Delta T = 0 $, we can use the following simplification;
\[G_{2\alpha}^{ ({\rm c})}(-E)=F_{ \nu 1}^{ ({\rm c})}( E)=F( E)\]
 On using above identity and applying delta function in one of the integral, Eq. (\ref{cot}) reduces to;
\begin{align}
\gamma_{\alpha \nu} ^{ ({\rm c})}
&=\frac{\hbar}{2\pi e^{4}R_{\alpha}R_{\nu}}\int d\epsilon F(-\epsilon) F(\epsilon+\Delta E)\left|\frac{1}{\epsilon+E_{1}-i\eta}-\frac{1}{\epsilon+E_{2}-i\eta}\right|^{2}
\end{align}
where, $E_{1}=E^{+}$; $E_{2}=-E^{-}+\Delta E$ and $\eta \rightarrow 0$ is applied to regularize the divergent integral in Eq. (\ref{cot}). All the co-tunneling rates involving either two islands or one island can be written in this form with corresponding $E_{1}$, $E_{2}$ and resistances involved. The regularization method,  described in ref. \cite{Kaasbjerg2016,Turek2002}, involves the removal of divergent terms using the sequential transition rates. \\
\ \\
Using $n(E)=1/\left(e^{E/k_{\rm B}T}-1\right)$, we get
\[F (-\epsilon)F(\epsilon+\Delta E)=-\epsilon \left(\epsilon+\Delta E\right)\left[n(\epsilon +\Delta E)-n(\epsilon)\right]n(\Delta E),\]
\[n(E)=\frac{1}{e^{E/k_{\rm B}T}-1}=-\frac{1}{2}\left[1-i\text{Cot} \left(\frac{iE}{2k_{\rm B}T}\right)\right].\]
Using the identity, \[\psi\left(1-z\right)-\psi\left(z\right)=\pi \text{Cot}(\pi z),\]
we obtain:
\begin{equation}
n(\epsilon+\Delta E)-n(\epsilon)=\frac{i}{2\pi}\left[\psi\left(1-\left(\frac{i\beta(\epsilon+\Delta E)}{2\pi}\right)\right)-\psi\left(1-\frac{i\beta \epsilon}{2\pi}\right)-\psi \left(\frac{i\beta (\epsilon +\Delta E)}{2\pi}\right)+\psi\left(\frac{i\beta \epsilon}{2\pi}\right)\right].
\end{equation}
Representing,
\[\psi^{-}(\epsilon')=\psi\left(1-\frac{i\beta \epsilon '}{2\pi}\right)\]
\[\psi^{+}(\epsilon ')=\psi\left(\frac{i\beta \epsilon'}{2\pi}\right),\]
we may write the co-tunneling rate in complex form as:
\begin{equation}
\gamma_{\alpha \nu}^{ ({\rm c})}=\kappa \int_{-\infty}^{\infty}dz ~g(z)\left[\psi^{-}(z+\Delta E)-\psi^{-}(z)+\psi^{+}(z)-\psi^{+}(z+\Delta E)\right]\left|\frac{1}{z+E_{1}-i\eta}-\frac{1}{z+E_{2}-i\eta}\right|^{2}
\label{cotc}
\end{equation}
where $g(z)=z(z+\Delta E)$ and $\kappa = \frac{-i\hbar n(\Delta E)}{4\pi^{2}e^{4}R_{\alpha}R_{\nu}}$.
With,
\begin{equation}
I^{-}(E_{1},E_{2})=\kappa \int_{-\infty}^{\infty}dz ~g(z)\Delta \psi^{-}(z)\frac{\left(E_{2}-E_{1}\right)^{2}}{\left[\left(z+E_{1}\right)^{2}+\eta^{2}\right]\left[\left(z+E_{2}\right)^{2}+\eta^{2}\right]}
\label{ei}
\end{equation}
and,
\begin{equation}
I^{+}(E_{1},E_{2})=\kappa \int_{-\infty}^{\infty}dz ~g(z)\Delta \psi^{+}(z)\frac{\left(E_{2}-E_{1}\right)^{2}}{\left[\left(z+E_{1}\right)^{2}+\eta^{2}\right]\left[\left(z+E_{2}\right)^{2}+\eta^{2}\right]}
\label{eij}
\end{equation}
where,
\[\Delta \psi ^{\pm}(-E_{i})=\psi^{\pm}(\Delta E-E_{i})-\psi^{\pm}(-E_{i})\]
the co-tunneling rate can be written in compact form as;
\begin{equation}
\gamma_{\alpha \nu}^{(c)}= I^{-}(E_{1},E_{2})-I^{+}(E_{1},E_{2}).
\end{equation}
\subsubsection*{1. Calculation of the Residues}
To solve the integral in Eq. (\ref{cotc}), we break it into $\psi^{+}$ and $\psi^{-}$ terms in Eq. (\ref{ei}) and Eq. (\ref{eij}) to have poles due to the digamma functions only either on upper half or lower half of complex plane. Now, to evaluate $I^{-}$, we close our contour in upper complex plane using an infinite radius semi-circle so that we have no poles from $\psi^{-}$ inside the contour. Hence, the only pole enclosed by the contour is given by;
\[z=-E_{i}+i\eta \]
The residue for above poles can be calculated to obtain;
\begin{equation}
a_{-1}^{(i)}(I^{-})=\frac{\alpha g(i\eta-E_{i})\Delta \psi^{-}\left(-E_{i}+i\eta\right)\left(E_{j}-E_{i}\right)}{2i\eta\left(E_{j}-E_{i}+2i\eta\right)}
\end{equation}
Similarly, to evaluate $I^{+}$, we close our contour in the lower complex plane so that we have no poles from $\psi^{+}$ inside the contour. Hence, the only pole enclosed by the contour is given by;
\[z=-E_{i}-i\eta\]
The residue for above pole can be calculated to obtain;
\begin{equation}
a_{-1}^{(i)}(I^{+})=\frac{\alpha g(i\eta-E_{i})\Delta \psi^{+}\left(-E_{i}-i\eta\right)\left(E_{j}-E_{i}\right)}{-2i\eta\left(E_{j}-E_{i}-2i\eta\right)}
\end{equation}
\subsubsection*{2. Calculation of $I(E_{1},E_{2})$}
The integral in Eq. (\ref{ei}) can be written as  sum of residues as;
\begin{equation}
I^{-}(E_{1},E_{2})=\frac{\pi\kappa}{\eta}\sum_{ij}\left[\frac{g\left(-E_{i}+i\eta\right)\Delta \psi^{-}\left(-E_{i}+i\eta\right)\left(E_{j}-E_{i}\right)}{\left(E_{j}-E_{i}+2i\eta\right)}\right]
\label{rei}
\end{equation}
We Taylor expand in $\eta$, the term inside the square bracket of Eq. (\ref{rei}). We keep only the first order term in $\eta$ (which eventually is independent of $\eta$ as observed from Eq. (\ref{rei})) and we remove the zeroth order term  which diverges when $\eta\rightarrow 0$. We get:
\begin{equation}
I^{-}\left(E_{1},E_{2}\right)=2\pi i\kappa \sum_{ij}\left[\frac{g(-E_{i})}{E_{i}-E_{j}}\Delta \psi^{-}(-E_{i})+\frac{1}{2}g'(-E_{i})\Delta \psi^{-}(-E_{i})-\frac{i\beta}{4\pi}g(-E_{i})\Delta \psi_{1}^{-}(-E_{i})\right]
\end{equation}
where;
\[\Delta \psi_{1} ^{\pm}(-E_{i})=\psi_{1}^{\pm}(\Delta E-E_{i})-\psi_{1}^{\pm}(-E_{i})\]
and $\psi_{1}$ represents the first derivative of $\psi$.
Next, we will solve integral (\ref{eij}) using similar approach. We obtain
\begin{equation}
I^{+}\left(E_{1},E_{2}\right)=-2\pi i\kappa \sum_{ij}\left[\frac{g(-E_{i})}{E_{i}-E_{j}}\Delta \psi^{+}(-E_{i})+\frac{1}{2}g'(-E_{i})\Delta \psi^{+}(-E_{i})-\frac{i\beta}{4\pi}g(-E_{i})\Delta \psi_{1}^{+}(-E_{i})\right]
\label{reij}
\end{equation}
\subsubsection*{3. Contribution from the semi-circle arcs}

To calculate the contribution form the semi-circle arcs, we consider the case for $z\rightarrow \infty$. We use following asymptotic expansion for the digamma function
\begin{equation}
\psi(z)\Big|_{z\rightarrow \infty}\approx \ln (z)-\frac{1}{2z}+O(z^{-2}).
\label{dig}
\end{equation}
Using Eq. (\ref{dig}), we obtain
\[\Delta \psi ^{+}(\epsilon)=\psi^{+}(\epsilon+\Delta E)-\psi^{+}(\epsilon)=\psi\left(\frac{i\beta}{2\pi}(\epsilon+\Delta E)\right)-\psi\left(\frac{i\beta}{2\pi}\epsilon\right)\approx\frac{\Delta E}{\epsilon} +O(\epsilon^{-2}) \]
\[\Delta \psi ^{-}(\epsilon)=\psi^{-}(\epsilon+\Delta E)-\psi^{-}(\epsilon)=\psi\left(1-\frac{i\beta}{2\pi}(\epsilon+\Delta E)\right)-\psi\left(1-\frac{i\beta}{2\pi}\epsilon\right)\approx\frac{\Delta E}{\epsilon} +O(\epsilon^{-2}) \]
and,
\[\left|\frac{1}{z+E_{1}-i\eta}-\frac{1}{z+E_{2}-i\eta}\right|_{z\rightarrow \infty}^{2}\sim \left|z\right|^{-4}.\]
By simple power counting, we find
\[\gamma_{\alpha \nu}^{\rm (c)}({\rm arc})\sim \int_{-\pi}^{\pi}d\theta R \cdot R^{-1} \cdot R^{-4}\cdot g(R)f(i\theta)\sim K R^{-4}g(R)\]
where, in this case $g(R)\sim R^{2}$ which implies the semi-circle arc does not contribute. In general, there is no contribution from the semi-circle arcs if $g(R)\sim R^{n}$ with $n<4$.
 \ \\
 \ \\
\subsubsection*{Final solution for the co-tunneling rate ($\Delta T = 0$ case)}
Including all the contributions the co-tunneling rate in the Eq. (\ref{cot}) can be written as;
\begin{equation}
\gamma_{\alpha \nu}^{\rm (c)}=2\pi i\kappa\sum_{i,j=1,2}\left[\frac{-i\beta}{4\pi}g(-E_{i})\left[\Delta \psi_{1}^{-}(-E_{i})-\Delta \psi_{1}^{+}(-E_{i})\right]+\left[\frac{1}{2}g'(-E_{i})+\frac{g(-E_{i})}{E_{i}-E_{j}}\right]\left[\Delta \psi ^{+}(-E_{i})+\Delta \psi ^{-}(-E_{i})\right]\right].
\label{fincot}
\end{equation}
\section*{2. Co-tunneling energy rates ($\Delta T = 0$ case)}
The co-tunneling energy rate for a process when an electron tunnels from $\alpha$ into island 2 and at the same time an electron leaves island 1 is given below. In this process, energy flows out of the reservoir $\alpha$.
\begin{align}
\gamma_{\alpha \nu}^{\rm (h)}(out)=&\frac{\hbar}{2\pi e^{4}R_{\alpha}R_{\nu}}\int d E_{\alpha}d E_{\nu}d E_{s_{1}}d E_{s_{2}}~{E_{\alpha}}~f_{\alpha}(E_{\alpha})\left(1-f_{\nu}(E_{\nu})\right)f(E_{s_{1}})\left(1-f(E_{s_{2}}\right)\nonumber \\
&\left|\frac{1}{E^{+}+E_{s_{2}}-E_{\alpha}}+\frac{1}{E^{-}+E_{\nu}-E_{s_{1}}}\right|^{2}\delta \left(E_{\nu}-E_{\alpha}+E_{s_{2}}-E_{s_{1}}+\Delta E\right)\nonumber \\
&=\frac{\hbar}{2\pi e^{4}R_{\alpha}R_{\nu}}\int  d \xi_{\alpha} d\xi_{\nu}F_{\alpha 2}^{(h)}(-\xi_{\alpha})G_{1 \nu}^{(c)}(\xi_{\nu})\left| \frac{1}{E^{+}+\xi_{\alpha}}+\frac{1}{E^{-}+\xi_{\nu}} \right|^{2}\delta\left(\xi_{\nu}+\xi_{\alpha}+\Delta E\right)
\label{ecotl}
\end{align}
But, when $\Delta T =0$, $F_{\alpha i}^{\rm (h)}(\Delta E)=\frac{1}{2}\Delta E \, F_{\alpha i}^{\rm (c)}(\Delta E)$. Doing some algebra, we obtain for the co-tunneling energy rates the same expression as for the co-tunneling charge rates ( see Eq. (\ref{fincot})) with the function `$g$' defined differently as,
\begin{equation}
\gamma_{\alpha \nu}^{\rm (h)}(out)=\gamma^{\rm (c)}_{\alpha \nu}\left[g(\epsilon)=-\frac{1}{2}\epsilon (\epsilon+\Delta E)\epsilon\right]
\label{ecott}
\end{equation}
Similarly, lets define the co-tunneling energy rate for the electrons entering into the reservoir $\alpha$;
\begin{align}
\gamma_{\alpha \nu}^{\rm (h)}(in)=&\frac{\hbar}{2\pi e^{4}R_{\alpha}R_{\nu}}\int d E_{\alpha}d E_{\nu}d E_{s_{1}}d E_{s_{2}}~~{E_{\alpha}}~f_{\nu}(E_{\nu})\left(1-f_{\alpha}(E_{\alpha})\right)f(E_{s_{2}})\left(1-f(E_{s_{1}}\right)\nonumber \\
&\left|\frac{1}{E^{+}+E_{s_{1}}-E_{\nu}}+\frac{1}{E^{-}+E_{\alpha}-E_{s_{2}}}\right|^{2}\delta \left(E_{\alpha}-E_{\nu}+E_{s_{1}}-E_{s_{2}}+\Delta E\right)\nonumber \\
\label{ecotr}
\end{align}
We obtain similar co-tunneling rates as in Eq.(\ref{ecott}) but with different expression for $g(\epsilon)$ given by
\begin{equation}
\gamma_{\alpha \nu}^{\rm (h)}(in)=\gamma_{\alpha \nu}^{\rm (c)}\left[g(\epsilon)=-\frac{1}{2}(\epsilon+\Delta E) (\epsilon+\Delta E)\epsilon\right]
\label{cott0in}
\end{equation}
All other co-tunneling energy rates can be written in the form of Eq. (\ref{ecott}) and Eq. (\ref{cott0in}) with suitable modification for energy parameters and resistances involved.
\subsection*{Co-tunneling rates $\Delta T \neq 0$}
In the presence of both thermal and potential bias, we cannot solve the integrals involved in the co-tunneling rates analytically. Although, we can still write the co-tunneling rates in the compact form  using Eq. (\ref{cotfun1}) and Eq. (\ref{cotfun}). In this section, we will suggest a proper regularization method for integrals in Eq. (\ref{cotfun1}) and Eq. (\ref{cotfun}) and simplify it to a form which can be easily integrated numerically.
We have;
\begin{equation}
H_{n_1,n_2}^{\rm{(c/h)}}\left(E_{1},E_{2}, E_{3}\right)=\frac{\hbar}{2\pi}\int_{-\infty}^{\infty} d\xi {F}_{\alpha 2}^{\rm{(c/h)}}(-\xi) F_{\nu 1}^{(c)}\left(\xi+ E_{3}\right)\left|\frac{1}{\xi+E_{1}-i\eta}-\frac{1}{\xi+E_{3}-E_{2}-i\eta}\right|^{2}
\end{equation}
Simplifying the term in square modulus, we obtain:
\begin{align}
H_{n_1,n_2}^{\rm{(c/h)}}\left(E_{1},E_{2}, E_{3}\right)&= \frac{\hbar}{2\pi}\bigg[\int d\xi ~\frac{{F}_{\alpha 2}^{\rm{(c/h)}}(-\xi) F_{\nu 1}^{(c)}\left(\xi+ E_{3}\right)}{\left(\xi +E_{1}\right)^{2}+\eta^{2}}+\int d\xi ~\frac{{F}_{\alpha 2}^{\rm{(c/h)}}(-\xi) F_{\nu 1}^{(c)}\left(\xi+ E_{3}\right)}{\left(\xi +E_{3}-E_2\right)^{2}+\eta^{2}}\nonumber \\
&-2\int d\xi ~{F}_{\alpha 2}^{\rm{(c/h)}}(-\xi) F_{\nu 1}^{(c)}\left(\xi+ E_{3}\right)\frac{\left[\left(\xi+E_{1}\right)\left(\xi+E_{3}-E_2\right)+\eta^{2}\right]}{\left[\left(\xi+E_{1}\right)^{2}+\eta^{2}\right]\left[\left(\xi+E_{3}-E_2\right)^{2}+\eta^{2}\right]}\bigg]
\label{com}
\end{align}
Now, lets transform the first two terms in Eq. (\ref{com}) such that $\xi \rightarrow z - E_1$ for the first term and $\xi \rightarrow z - E_3 + E_2$ for the second term.
\begin{align}
H_{n_1,n_2}^{\rm{(c/h)}}\left(E_{1},E_{2}, E_{3}\right)&= \frac{\hbar}{2\pi}\bigg[\int dz ~\frac{{F}_{\alpha 2}^{\rm{(c/h)}}(-z+E_1) F_{\nu 1}^{(c)}\left(z+E_3-E_1 \right)}{z^{2}+\eta^{2}}+\int d\xi ~\frac{{F}_{\alpha 2}^{\rm{(c/h)}}(-z+E_3-E_2) F_{\nu 1}^{(c)}\left(z+ E_{2}\right)}{z^{2}+\eta^{2}}\nonumber \\
&-2\int d\xi ~{F}_{\alpha 2}^{\rm{(c/h)}}(-\xi) F_{\nu 1}^{(c)}\left(\xi+ E_{3}\right)\frac{\left[\left(\xi+E_{1}\right)\left(\xi+E_{3}-E_2\right)+\eta^{2}\right]}{\left[\left(\xi+E_{1}\right)^{2}+\eta^{2}\right]\left[\left(\xi+E_{3}-E_2\right)^{2}+\eta^{2}\right]}\bigg]
\end{align}
We use the approach in reference \cite{Turek2002} to regularize the integral, i.e.
\[\lim_{\eta\rightarrow 0}\int dz ~\frac{g(z-E_{i})}{z^{2}+\eta^{2}}\longrightarrow \int dz ~\frac{g(z-E_{i})-g(-E_{i})}{z^{2}}.\]
So, when $\eta \rightarrow 0$,
\begin{align}
H_{n_1,n_2}^{\rm{(c/h)}}\left(E_{1},E_{2}, E_{3}\right)&= \frac{\hbar}{2\pi}\bigg[\int dz ~\frac{{F}_{\alpha 2}^{\rm{(c/h)}}(-z+E_1) F_{\nu 1}^{(c)}\left(z+E_3-E_1 \right)-{F}_{\alpha 2}^{\rm{(c/h)}}(E_1) F_{\nu 1}^{(c)}\left(E_3-E_1 \right)}{z^{2}}\nonumber \\
&+\int dz ~\frac{{F}_{\alpha 2}^{\rm{(c/h)}}(-z+E_3-E_2) F_{\nu 1}^{(c)}\left(z+ E_{2}\right)-{F}_{\alpha 2}^{\rm{(c/h)}}(E_3-E_2) F_{\nu 1}^{(c)}\left( E_{2}\right)}{z^{2}}\nonumber \\
&-2\int d\xi ~{F}_{\alpha 2}^{\rm{(c/h)}}(-\xi) F_{\nu 1}^{(c)}\left(\xi+ E_{3}\right)\frac{\left[\left(\xi+E_{1}\right)\left(\xi+E_{3}-E_2\right)\right]}{\left[\left(\xi+E_{1}\right)^{2}\right]\left[\left(\xi+E_{3}-E_2\right)^{2}\right]}\bigg]
\label{commm}
\end{align}
The integrals in Eq. (\ref{commm}) are properly regularized and can be numerically evaluated for the case of both thermal and potential bias.
\section{Charge and heat current in the co-tunneling regime}
\label{Curr}
The expression for charge and heat currents flowing towards the right reservoir in contact with island 2 is given by
\begin{align}
{I}_{\rm R{2}}^{\rm (c/h)}&=Q^{(c/h)}\Bigg[\Big[\gamma_{\rm L{1}R{2}}^{\rm(c/h)}(n_{1},n_{2}+1)+\gamma_{\rm R{1}R{2}}^{(c/h)}(n_{1},n_{2}+1)+\Gamma_{2,\rm R {2}}^{\rm (c/h)}(n_{1},n_{2})\Big]p(n_{1},n_{2}+1)\nonumber \\
&+\Big[\gamma_{\rm L{1}R{2}}^{\rm (c/h)}(n_{1}+1,n_{2}+1)+\gamma_{\rm R{1}R{2}}^{\rm (c/h)}(n_{1}+1,n_{2}+1)+\Gamma_{2, \rm R{2}}^{\rm (c/h)}(n_{1}+1,n_{2})\Big]p(n_{1}+1,n_{2}+1)\nonumber \\
&-\Big[\gamma_{\rm L{1}R{2}}^{\rm (c/h)}(n_{1},n_{2})+\gamma_{\rm R{1}R{2}}^{\rm (c/h)}(n_{1},n_{2})+\Gamma_{\rm R{2},2}^{\rm (c/h)}(n_{1},n_{2})\Big]p(n_{1},n_{2})\nonumber \\
&-\Big[\gamma_{\rm L {1}{\rm R}{2}}^{\rm(c/h)}(n_{1}+1,n_{2})+\gamma_{\rm R{1}\rm R{2}}^{\rm (c/h)}(n_{1}+1,n_{2})+\Gamma_{\rm R{2}, 2}^{\rm (c/h)}(n_{1}+1,n_{2})\Big]p(n_{1}+1,n_{2})\Bigg] ,
\label{current}
\end{align}
where $Q^{\rm (c)} = e$ and $Q^{(h)}=1$.
${I}_{\rm L{2}}^{\rm (c/h)}$ can be written anagously.
Eq.~(\ref{current}) can be broken down into the one containing only sequential tunneling rates and another one containing only the co-tunneling rates.
\end{widetext}

\section{Time-dependent drag heat current in quantum wires}
\label{QWtd}
\begin{figure}[t]
\centering
\includegraphics[width=1\columnwidth,clip=true]{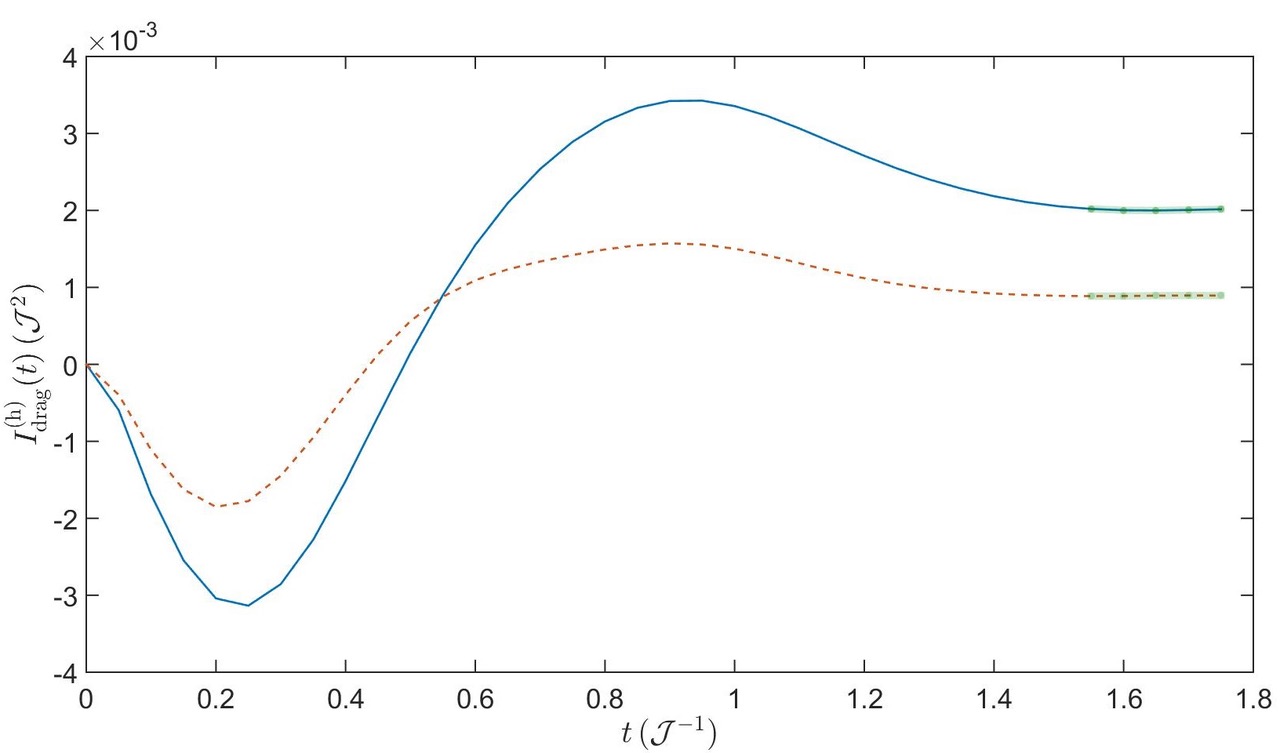}
\caption{Plot of the time-dependent $I^{\rm (h)}_{\rm{drag}}(t)$ calculated for $\beta_L=0.5\mathcal{J}^{-1}$, $\beta_R=0.75\mathcal{J}^{-1}$, $\Delta=0$ and $U=0.3\mathcal{J}$ (solid line) or $U=0.2\mathcal{J}$ (dashed line). The stationary state values of $I^{\rm h}_{\rm{drag}}$ are marked in green.}
\label{J2t}
\end{figure}

The time-dependence of the drag current $I^{\rm (h)}_{\rm drag}$, in the case of two parallel quantum wires, is numerically simulated by using the protocol detailed in Sec.~\ref{numsim}.  The resulting drag current $I^{\rm (h)}_{\rm drag}$ is plotted as a function of time in Fig.~\ref{J2t} for $\Delta=0$ and two values of inter-wire coupling, namely $U=0.3\mathcal{J}$ and $U=0.2\mathcal{J}$, respectively.
We observe that the time-dependent current presents a negative peak, followed by a positive peak thereafter reaching the stationary plateau value. The transient time (which is of order $\sim\cal J$) loosely depend on $\Delta$, but does not appear to depend on $U$ or on the temperature.
We are not interested in this transient behavior, though, which actually depends on the initial conditions.
We instead focus on the stationary value of $I^{\rm (h)}_{\rm drag}$, which we calculate by taking the average of the current over the plateau, marked in green in Fig.~\ref{J2t}.


\begin{thebibliography}{30}
\bibitem{nota_endep}
Lead-island couplings are energy-dependent when their tunnelling matrix elements depend on energy or when at least one density of states (of the lead or of the island) depend on energy.
\bibitem{heat_con}
The energy current, defined as $I^{\rm (e)}_{\alpha i}=I^{\rm (h)}_{\alpha i}+eV_{\alpha i}\, I^{\rm (c)}_{\alpha i}$, is not conserved even globally (apart from specific values of the temperatures of the islands).
Indeed, fixing the temperature of the islands is equivalent to connect the islands to reservoirs, thus allowing them to exchange energy with the ``environment".
\bibitem{footnote}
Note that the cross-term $V\Delta T$ vanishes when $n_{x1}=n_{x2}=1/2$.
\bibitem{ratesnx2}
For $n_{x2}>1/2$ the two chemical potentials shift downwards, thus reducing the rate for exiting the island.
For $n_{x2}<1/2$ the two chemical potentials shift upwards, thus reducing the rate for entering the island.
\end{thebibliography}
\end{document}